\documentclass[aps,twocolumn,raggedbottom,prb,showpacs,nobalancelastpage,amssymb,groupedaddress]{revtex4}
\usepackage{graphicx}
\usepackage{amsmath}
\usepackage{amssymb}
\newcommand{\bra}[1]{\langle #1|}	%
\newcommand{\ket}[1]{|#1\rangle}
\newcommand{\braket}[2]{\langle #1|#2\rangle}
\usepackage{dsfont}

\begin{document}
\title{Dissipative dynamics of a qubit coupled to a nonlinear oscillator}
\author{Carmen\, Vierheilig, Johannes\, Hausinger, and Milena\, Grifoni}
\affiliation{Institut f\"{u}r Theoretische Physik, Universit\"at
Regensburg, 93035 Regensburg, Germany}
\date{\today}

\begin{abstract}
We consider the dissipative dynamics of a qubit coupled to a nonlinear oscillator (NO) embedded in an Ohmic environment. By treating the nonlinearity up to first order and applying Van Vleck perturbation theory up to second order in the qubit-NO coupling, we derive an analytical expression for the eigenstates and eigenfunctions of the coupled qubit-NO system beyond the rotating wave approximation. In the regime of weak coupling to the thermal bath, analytical expressions for the time evolution of the qubit's populations are derived: they describe a multiplicity of damped oscillations superposed to a complex relaxation part toward thermal equilibrium. The long-time dynamics is characterized by a single relaxation rate, which is maximal when the qubit is tuned to one of the resonances with the nonlinear oscillator.
\end{abstract}
 \maketitle

\section{Introduction}
Coupling a two-level system (TLS) to a harmonic oscillator has attracted a lot of attention in various fields of physics. Examples are two-level quantum dots in photonic crystal nanocavities \cite{Yoshie,Faraon}, a quantum dot exciton in a microcavity \cite{Reithmaier}, or single atoms with a large dipole moment interacting with photons in a microwave cavity \cite{Raimond}. Within the framework of quantum computation two prominent solid-state realizations of a qubit-oscillator system are found: a Cooper-pair box \cite{Nakamura1999,Makhlin,Vion2002,Collin2004} coupled to a transmission line resonator \cite{Blais,Wallraff,Schuster2005,Wallraff2005,SchusterResolving,WallraffSideband} and the Josephson flux qubit \cite{Mooij1999} read-out by a DC-SQUID\cite{Chiorescu2003,Chiorescu2004,Johansson2006}. 
The Cooper-pair box setup has been used to perform non-demolition measurements or to transfer information between qubits via the transmission line resonator \cite{Blais,Blais2007,Wallraff2005,Houck,Bishop,Fink}.
 In the second experimental realization the flux qubit is usually read-out via a damped DC-SQUID, which acts as a linear or nonlinear oscillator. 
A non-demolition read-out procedure, based on the measurement of the Josephson inductance, is given by Lupa\c{s}cu et al. \cite{Lupascu2004}.\\
At present the effort to exploit the nonlinearity of a qubit read-out device, for example, a DC-SQUID or a Josephson bifurcation amplifier (JBA) \cite{Siddiqi2,Siddiqi}, is growing, as nonlinear effects lead to advantages in various measurement schemes and to new physical observations. For example, the qubit read-out can be optimized by using the SQUID in the nonlinear regime as a bifurcation amplifier leading to fast read-out with high fidelity \cite{Lee,Picot}. Second, the bifurcation allows for a higher sensitivity when determining the qubit states and, due to the nonlinear Josephson inductance, a high quality factor for the resonance is achieved \cite{Siddiqi}. However, the nonlinear regime also provides new channels of relaxation \cite{Picot}. %
Moreover there are recent experiments embedding a micromechanical resonator in a nonlinear DC-SQUID, which is strongly damped to avoid bistability, to acquire cooling and squeezing of the resonator modes and to achieve quantum-limited position detection \cite{Etaki}. Such a composed system can then also be coupled to a qubit. Besides these examples a SQUID which is embedded into a cavity \cite{Nation} can be used as a bifurcation amplifier in its nonlinear regime.\\
All these approaches rely in principle on treating the SQUID as a classical nonlinear system. To our knowledge there has been to date no experimental realization of a SQUID in the nonlinear quantum regime.\\
From the theoretical point of view nonlinear quantum oscillators have been predominantly studied within the context of the quantum Duffing oscillator model \cite{Fistul,Peano1,Peano2,Peano3,Serban}, where the oscillator is subject to an external ac driving force.
Strikingly, the response of the Duffing oscillator displays antiresonant dips and resonant peaks depending on the frequency of the driving field \cite{Fistul}. The antiresonances persist in the presence of a weak Ohmic bath; for high damping the nonlinear response of the oscillator resembles the one of a linear oscillator at a shifted frequency \cite{Peano1,Peano2,Peano3}.\\
Despite the numerous theoretical works on coupled qubit-linear oscillator systems \cite{Garg,Tian,vanderWal,Thorwart,Goorden2004,Goorden2005,Wilhelm1,Kleff1,Kleff2} the case of a TLS-Josephson bifurcation amplifier system has been addressed only very recently by Nakano et al. \cite{Nakano}.
Here we study the SQUID as a nonlinear, undriven oscillator acting as a read-out device for a qubit. We consider weak nonlinearities  such that the corresponding linear system can be retained at any step of our calculation. With the help of Van Vleck perturbation theory in the TLS-oscillator coupling $g$ we determine the eigenstates and spectrum of the coupled system and the corresponding dynamics in analytic form. Thus we can quantitatively characterize the influence of
 the coupling $g$ and of the nonlinearity on the dynamics of the composed system. The overall effects of the nonlinearity are the following: (i) a shift of the transition frequencies to higher values compared to the linear case; (ii) the amplitudes associated to the transition frequencies are modified. In particular the vacuum Rabi splitting is decreased by the interplay of coupling and nonlinearity. To account for dissipative effects we add a weak Ohmic environment.
Then the dynamics of the reduced density matrix of the composed system can be described in terms of a set of coupled differential equations for its matrix elements in the energy basis (Bloch-Redfield equations). We discuss a partial secular approximation (PSA) to those equations as well as two more stringent approximations, the full secular approximation in the low temperature approximation (LTA) and the smallest eigenvalue approximation (SEA) accounting for the long time dynamics. All these three approximation schemes allow for analytical solution  of the dynamics of the TLS, which we compare with predictions obtained by numerically solving the Bloch-Redfield equations. It turns out that the most accurate PSA should be used when investigating strong nonlinearities. The long-time approximation enables us nevertheless to extract the correct relaxation rate within the regime of validity of our perturbative approach.
The paper is organized as follows: In section \ref{model} we introduce the model with the relevant dynamical quantities. In section \ref{secEnspec} the energy spectrum and the dynamics of the non-dissipative coupled system is investigated. Section \ref{env} addresses the dissipative effects, while in section \ref{secRes} results are represented. In section \ref{Conclusions} conclusions are drawn.

\section{The model}\label{model}
\subsection{Qubit-nonlinear oscillator-bath system}
In this section we consider a TLS coupled to a nonlinear oscillator, which itself is coupled to an Ohmic bath. This model mimics, e.g., the situation of a flux qubit, made of three Josephson junctions, which is coupled inductively to a damped DC-SQUID \cite{Lee,Picot}. The qubit with its two logical states, the clockwise and counterclockwise currents, represents a two-level system. Because the SQUID itself is coupled to an environment, it transfers environmental influences which lead to the dissipation in the qubit. 
Hence the total Hamiltonian reads:
\begin{eqnarray}\label{Htot}
\mathcal{H}=\mathcal{H}_{\rm TLS-NO}+\mathcal{H}_{\rm NO-B}+\mathcal{H}_{\rm B},
\end{eqnarray}
with $\mathcal{H}_{\rm TLS-NO}$ describing the coupled TLS-nonlinear oscillator system, while $\mathcal{H}_{\rm NO-B}$ and $\mathcal{H}_{\rm B}$ are the coupling between the oscillator and bath and the bath Hamiltonian, respectively. For later convenience we write %
\begin{eqnarray}
\mathcal{H}_{\rm TLS-NO}=\underbrace{\mathcal{H}_{\rm TLS}+\mathcal{H}_{\rm NO}}_{\mathcal{H}_{0}}+\mathcal{H}_{\rm Int}
\end{eqnarray}
with coupling Hamiltonian $\mathcal{H}_{\rm Int}$.

\subsubsection{Two-level system}
First we consider the Hamiltonian of the TLS,
\begin{eqnarray}
\mathcal{H}_{\rm TLS}&=&-\frac{\hbar}{2}\left(\varepsilon\sigma_z+\Delta_0\sigma_x\right),
\end{eqnarray}
represented in the localized basis $\left\{|L\rangle,|R\rangle\right\}$ \cite{Weiss}, corresponding to clockwise and counterclockwise currents, respectively, in the superconducting ring. The $\sigma_{i}$, $i=x,z$, are the corresponding Pauli matrices. The energy bias $\varepsilon$ can be tuned for a superconducting flux qubit by application of an external flux $\Phi_{\rm ext}$ and vanishes at the so-called degeneracy point \cite{WallraffSideband}.\\
For $\varepsilon\gg\Delta_0$, where $\Delta_0$ is the tunneling amplitude, the states $|L\rangle$ and $|R\rangle$ are eigenstates of the TLS, while at the degeneracy point the eigenstates $\ket{\rm g}$, $\ket{\rm e}$ are given by symmetric and antisymmetric superpositions, respectively, of the two logical states. %
 In general the states $|R\rangle$ and $|L\rangle$ become in the energy basis:
\begin{eqnarray}
    \ket{R} &= \cos (\Theta/2) \ket{\rm g} +\sin (\Theta/2) \ket{\rm e} , \label{LocaltoEn1}\\
   \ket{L} &= -\sin (\Theta/2) \ket{\rm g} +\cos (\Theta/2) \ket{\rm e},  \label{LocaltoEn2}\nonumber
\end{eqnarray}
with $\tan \Theta  = - \Delta_0 / \varepsilon$ and $-\frac{\pi}{2} \leq \Theta < \frac{\pi}{2}$. Moreover in this basis the TLS Hamiltonian is: $\tilde{\mathcal{H}}_{\rm TLS} = -\frac{\hbar \Delta_b}{2} \tilde \sigma_{\rm z}$, where $\tilde{\sigma}_z$ is the Pauli matrix in the energy basis and  $\hbar\Delta_b=\hbar\sqrt{\varepsilon^2+\Delta_0^2}$ is the energy splitting.
\subsubsection{Nonlinear oscillator}
The Hamiltonian for the nonlinear oscillator is composed of a linear harmonic oscillator modified with a quartic term in the position operator,
\begin{eqnarray}\label{hno}
\mathcal{H}_{\rm NO}&=&\hbar\Omega \hat{j}+\frac{\alpha}{4}(B+B^\dagger)^4,
\end{eqnarray}
where $\hat{j}=B^\dagger B$ is the occupation number operator of the linear oscillator and $B$ and $B^\dagger$ are the corresponding annihilation and creation operators. 
In the following we restrict to the case of hard nonlinearities, i.e., $\alpha>0$.
Using time-independent perturbation theory we consider small nonlinearities $\alpha\ll\hbar\Omega$ and evaluate the eigenvalues $\mathcal{E}_j$ and eigenfunctions $\ket{j}$ of (\ref{hno}) to lowest order in the nonlinearity,
\begin{eqnarray}\label{enpert}
\mathcal{E}_j&:=&%
\hbar\Omega j+\frac{3}{2}\alpha j(j+1),\quad j=0,\ldots,\infty
\end{eqnarray}
\begin{eqnarray}\label{gl9}
|j\rangle&:=&|j\rangle_0+a_{-2}^{(j)}|j-2\rangle_0+a_{2}^{(j)}|j+2\rangle_0+\\
&&a_{-4}^{(j)}|j-4\rangle_0+a_{4}^{(j)}|j+4\rangle_0\nonumber,
\end{eqnarray}
where $|\rangle_0$ denotes the eigenstate of the corresponding linear oscillator. The expansion coefficients for the $j$th state of the nonlinear oscillator are given by:
\begin{eqnarray}
a_{-4}^{(j)}&=&\frac{\sqrt{(j-3) (j-2) (j-1) j}  \alpha }{16 \hbar \Omega},\\
a_{4}^{(j)}&=&-\frac{\sqrt{(j+1) (j+2) (j+3) (j+4)} \alpha }{16 \hbar \Omega}\nonumber,\\
a_{-2}^{(j)}&=&\frac{\left(j-\frac{1}{2}\right) \sqrt{(j-1) j} \alpha }{2 \hbar \Omega}\nonumber,\\
a_{2}^{(j)}&=&-\frac{\left(j+\frac{3}{2}\right) \sqrt{(j+1) (j+2)} \alpha }{2  \hbar\Omega}\nonumber.
\end{eqnarray}
We notice that two arbitrary eigenstates $\ket{j}$, $\ket{k}$ are orthonormal up to first order in the nonlinearity.\\
Perturbation theory for a nonlinear oscillator has to be elaborated carefully. Due to the special form of the nonlinear term, proportional to $(B+B^\dagger)^4$ the energy corrections acquire a strong level dependence: $\mathcal{E}_j^{(1)}=\frac{3}{2}\alpha j(j+1)$ for the first, see Eq. (\ref{enpert}), and $\mathcal{E}_j^{(2)}=\frac{1}{8\hbar\Omega}\alpha^2 (-34j^3-51j^2-59j-21)$ for the second order. Depending on the actual level number the second order can be as large as the first order for fixed nonlinearity. To avoid this, one has to choose the nonlinearity parameter $\alpha$ such that the oscillator levels  under consideration are well represented by the first order result. The error done by disregarding the $n$th order perturbation theory is estimated in the following by introducing $Er^{(n)}(j)=|\mathcal{E}_j^{(n)}|/\mathcal{E}_j^{(0)}$ for different nonlinearities (see Table \ref{table}). 
\begin{table}
\begin{tabular}{|c|c|c|c|}
\hline
Error&$\alpha/\hbar\Omega=10^{-3}$&$\alpha/\hbar\Omega=0.01$&$\alpha/\hbar\Omega=0.02$\\
\hline\hline
$Er^{(1)}(1)$&$3\cdot 10^{-3}$&$0.03$&$0.06$\\
$Er^{(2)}(1)$&$2.06\cdot 10^{-5}$&$2.06\cdot 10^{-3} $&$8.25\cdot 10^{-3} $ \\
$Er^{(1)}(2)$&$4.5
\cdot 10^{-3}$&$0.045$&$0.09$\\
$Er^{(2)}(2)$&$3.84\cdot 10^{-5}$&$3.8\cdot 10^{-3}$& $ 0.015$\\
$Er^{(1)}(3)$&$6\cdot 10^{-3}$&$0.06$&$0.12$\\
$Er^{(2)}(3)$&$6.56\cdot 10^{-5}$&$6.56\cdot 10^{-3}$&$0.026 $ \\
$Er^{(1)}(4)$&$7.5\cdot 10^{-3}$&$0.075$&$0.15 $ \\
$Er^{(2)}(4)$&$1.02\cdot 10^{-4}$&$1.02\cdot 10^{-2}$&$0.041$ \\
$Er^{(1)}(5)$&$9\cdot 10^{-3}$&$9\cdot 10^{-2}$&$ 0.18$ \\
$Er^{(2)}(5)$&$1.46\cdot 10^{-4}$&$1.46\cdot 10^{-2}$&$0.058$ \\
\hline
\end{tabular}
\caption{Error estimation for different values of the nonlinearity for the six lowest levels.\label{table}}
\end{table} 
Taking only first order perturbation theory into account, the error is determined by $Er^{(2)}(j_{\rm{max}})$, where $j_{\rm{max}}$ is the highest level under consideration. The error made by using first order perturbation theory is in case of $\alpha/\hbar\Omega=0.02$ around $6\%$ for the $j=5$ level.\\
Finally we consider a coupling Hamiltonian of the form:
\begin{eqnarray}
\mathcal{H}_{\rm Int}&=&\hbar g\sigma_z(B+B^\dagger).%
\end{eqnarray}
This kind of coupling arises due to the inductive coupling of the TLS to the SQUID \cite{Burkard}.
\subsubsection{Harmonic bath}
Following Caldeira and Leggett \cite{CaldeiraLeggett}, we model the environmental influences originating from the circuitry surrounding the qubit and the oscillator as a bath of harmonic oscillators being coupled bilinearly to the nonlinear oscillator. Thus, the environment is described by
  $\mathcal{H}_{\rm B} = \sum_k \hbar \omega_k b^\dagger_k b_k$
and the interaction Hamiltonian is
\begin{equation} \label{CouplingHam}
  \mathcal{H}_{\rm NO-B} = (B^\dagger + B) \sum_k \hbar \nu_k (b_k^\dagger + b_k) +  (B^\dagger + B)^2 \sum_k \hbar \frac{\nu_k^2}{\omega_k}.
\end{equation}
The operators $b_k^\dagger$ and $b_k$ are the creation and annihilation operators, respectively, for the $k$th bath oscillator, $\omega_k$ is its frequency, and $\nu_k$ gives the coupling strength. The whole bath can be described by its spectral density, which we consider to be Ohmic,
\begin{equation} \label{OhmicSpecDens}
  G_{ \rm{Ohm}} (\omega) = \sum_k  \nu_k^2 \delta (\omega - \omega_k) = \kappa \omega, 
\end{equation} 
where $\kappa$ is a dimensionless coupling strength.
\subsection{Population difference}\label{popdiff}
We wish to describe the dynamics $P(t)$ of the TLS described by the population difference 
\begin{eqnarray} \label{dynamics}
  P(t) &=&{ \rm} {\rm Tr}_{{\rm TLS}} \{ \sigma_{ {\rm z}} \rho_{{\rm red}} (t) \}\\& =& \bra{{\rm R}}  \rho_{{\rm red}}(t) \ket{{\rm R}} - \bra{{\rm L}}  \rho_{{\rm red}}(t) \ket{{\rm L}}\nonumber
\end{eqnarray}
between the $\ket{R}$ and $\ket{L}$ states of the qubit.
The reduced density matrix of the TLS,
\begin{equation}
  \rho_{{\rm red}}(t) ={ \rm Tr}_{{\rm{NO}}} {\rm Tr}_{{\rm B}} \{ W(t) \}= {\rm Tr}_{{\rm{NO}}} \{  \rho (t)\},
\end{equation} 
is found after tracing out the oscillator and bath degrees of freedom from the total density matrix
  $W (t) = \exp^{- \frac{{\rm i}}{\hbar} \mathcal{H} t} W (0) \exp^{\frac{{\rm i}}{\hbar}  \mathcal {H} t}$. For vanishing nonlinearities it is possible to map the problem  described by the Hamiltonian in equation (\ref{Htot}) onto a spin-boson model \cite{Garg} with an effective peaked spectral density depending on the coupling $g$, the frequency $\Omega$, and the damping strength $\kappa$. This mapping hence allows the evaluation of the population difference $P(t)$ of the TLS using standard approximations developed for the spin-boson model \cite{Goorden2004,Goorden2005,Nesi}.
 This mapping, however, does no longer hold true in the nonlinear oscillator case. 
Hence in this work we consider the TLS and the nonlinear oscillator as central quantum system and describe dissipative effects by solving the Bloch-Redfield master equations for the reduced density matrix $\rho(t) ={ \rm Tr}_{\rm B} \{ W(t) \}$ of the qubit-NO system. In a second step we perform the trace over the NO degrees of freedom to obtain the reduced dynamics of the TLS. An expression for $P(t)$ is then given in terms of diagonal and off-diagonal elements of $\rho(t)$ in the $\mathcal{H}_{\rm TLS-NO}$ Hamiltonian's eigenbasis $\{\ket{n} \}$. It reads \cite{Hannes}:
\begin{equation} \label{PtQubitHO}
  P(t) = \sum_n p_{nn}(t) + \sum_{\stackrel{n,m}{n > m}} p_{nm}(t),
\end{equation}  
where
\begin{eqnarray}\label{gl6}
  p_{nn}(t) &=& \sum_j \left\{ \cos \Theta \biggl[ \braket{j {\rm g}}{n}^2 - \braket{j {\rm e}}{n}^2 \biggr]  +\right.\\&&\left. 2 \sin \Theta \braket{j {\rm g}}{n} \braket{j {\rm e}}{n} \right\} \rho_{n n}(t) \label{p0},\nonumber\\
 p_{nm}(t) &=& 2 \sum_j \biggl\{ \cos \Theta \biggl[\braket{j {\rm g}}{n} \braket{m}{j {\rm g}} - \braket{j {\rm e}}{n} \braket{m}{j {\rm e}}    \biggr] \nonumber\\
&&     + \sin \Theta \biggl[\braket{j {\rm e}}{n} \braket{m}{j {\rm g}} + \braket{j {\rm e}}{m} \braket{n}{j {\rm g}}    \biggr] \biggr\}\nonumber\\&& {\rm Re} \{ \rho_{nm}(t) \} \label{pnm}\nonumber,
\end{eqnarray}
and $\rho_{nm}(t) = \bra{n} \rho(t) \ket{m}$. The TLS-NO eigenstates are derived in the next section.

\section{Energy spectrum and dynamics of the non-dissipative TLS-NO system}\label{secEnspec}
In the following we derive the eigenenergies and eigenstates of the unperturbed TLS-NO Hamiltonian $\mathcal{H}_{\rm TLS-NO}$ using Van Vleck perturbation theory \cite{Shavitt1980,Cohen1992}. This approach allows us to deal with spectra containing almost exactly degenerate levels organized in manifolds (here doublets), as it is the case if the TLS and nonlinear oscillator are close to resonance, $\Delta_b\approx\Omega$, and the coupling $g$ is small compared to the energy separation of the manifolds.

\subsection{Energy spectrum}
\begin{figure}[h]
\begin{flushright}
  \resizebox{\linewidth}{!}{
  \includegraphics{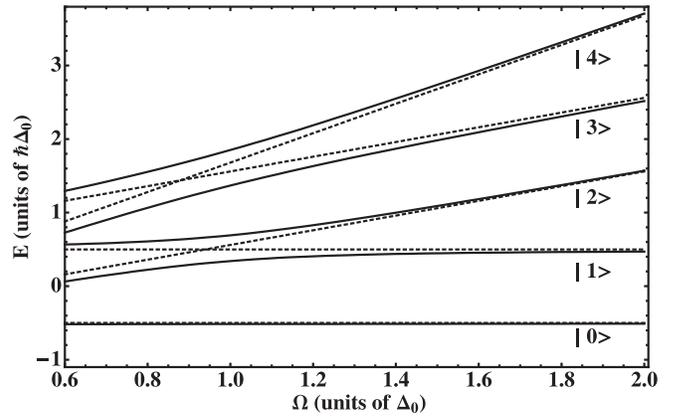}}
\end{flushright}
\caption{Energy spectrum of the coupled qubit-nonlinear-oscillator system versus the linear oscillator frequency $\Omega$ (in units of the TLS tunneling splitting $\Delta_0$). Solid lines show the energy levels for the five lowest energy states ($\ket{0}$, $\ket{1}$, $\ket{2}$, $\ket{3}$, $\ket{4}$) with the TLS-NO coupling being switched on, $g=0.18\Delta_0$, and for finite nonlinearity, $\alpha=0.02\hbar\Delta_0$. The TLS is unbiased, $\varepsilon=0$. The energy levels for the uncoupled case are given by the dotted lines.
Due to the non-equidistant level spacing of the nonlinear oscillator the resonance condition (crossing of dotted lines), given in equation (\ref{rescond}), is different for each doublet. This causes a shift of the exact crossings with respect to the linear case at zero coupling to lower frequencies. For finite coupling the spectrum exhibits avoided crossings around resonance, whereas it approaches the uncoupled case away from resonance.\label{EnSpec}}
\end{figure}

The eigenstates of the uncoupled TLS-NO system Hamiltonian $\tilde{\mathcal{H}}_0$ are $ \{ \ket{ j} \otimes \ket{{\rm g}}; \ket{j} \otimes \ket{{\rm e}} \} \equiv \{ \ket{j {\rm g}} ; \ket{j {\rm e}}  \}$.
The associated energies are depicted by the dotted lines in figure \ref{EnSpec}. 
At the resonance condition of the TLS with two neighboring nonlinear oscillator levels, 
\begin{eqnarray}\label{rescond}
\hbar\Omega&=&\hbar\Delta_b-3\alpha(j+1),
\end{eqnarray}
where $j$ denotes the lower oscillator level involved, the states $|(j+1){\rm g}\rangle$ and $|j{\rm e}\rangle$ are exactly degenerate except for the ground state $|0\rm g\rangle$. For finite coupling the full Hamiltonian $\mathcal{H}_{\rm TLS-NO}$ acquires in the basis $\{|j{\rm g}\rangle;|j{\rm e}\rangle\}$ the form
\begin{eqnarray}\label{gl30a}
\tilde{\mathcal{H}}_{\rm TLS-NO}&=&\tilde{\mathcal{H}}_0+\tilde{\mathcal{H}}_{\rm Int}\\
&=&-\frac{\hbar\Delta_b}{2}\tilde{\sigma}_z+\hbar \Omega \hat{j}+\frac{3}{2}\alpha \hat{j}(\hat{j}+1)+\nonumber\\&&\frac{\hbar g}{\Delta_b}\left(\epsilon\tilde{\sigma}_z-\Delta_0\tilde{\sigma}_x\right)\left(B+B^\dagger\right)\nonumber.
\end{eqnarray}
To find the eigenvalues of the Hamiltonian $\tilde{\mathcal{H}}_{\rm TLS-NO}$, we treat $\tilde{\mathcal{H}}_{\rm Int}\propto g$ as a small perturbation, which is satisfied for $g\ll\Delta_b,\Omega$. Using Van Vleck perturbation theory \cite{Shavitt1980,Cohen1992} we can construct an effective Hamiltonian by applying an unitary transformation to $\tilde{\mathcal{H}}_{\rm TLS-NO}$,
\begin{eqnarray}
\tilde{\mathcal{H}}_{\rm eff}&=&\exp(iS)\tilde{\mathcal{H}}_{\rm TLS-NO}\exp(-iS).
\end{eqnarray}
$\tilde{\mathcal{H}}_{\rm eff}$ has the same eigenvalues as $\tilde{\mathcal{H}}_{\rm TLS-NO}$ but does not involve matrix elements connecting states which are far away from degeneracy. Consequently it is block-diagonal with all quasi-degenerate energy levels being in one common block. Because the quasi-degenerate states form doublets, each block of $\tilde{\mathcal{H}}_{{\rm eff}}$ is given by a $2\times 2$ matrix. The latter can be diagonalized easily. To calculate $S$ and $\tilde{\mathcal{H}}_{{\rm eff}}$ we write both as a power series up to first order in the nonlinearity $\alpha$ and up to second order in the coupling $g$,
\begin{eqnarray}\label{q32}
S&=&S^{(0)}+S^{(1)}+S^{(2)}+\mathcal{O}(\alpha^2,g^3),\\
\tilde{\mathcal{H}}_{\rm eff}&=&\tilde{\mathcal{H}}_{\rm eff}^{(0)}+\tilde{\mathcal{H}}_{\rm eff}^{(1)}+\tilde{\mathcal{H}}_{\rm eff}^{(2)}+\mathcal{O}(\alpha^2,g^3),
\end{eqnarray}
where $\exp(iS^{(0)})=\mathds{1}$. The upper index in the above equation denotes the actual order in $g$. Consequently in the following we assume that $\alpha/\hbar\Omega\sim g^2/\Omega^2\ll1$. To calculate $S^{(1/2)}$ and $\tilde{\mathcal{H}}_{{\rm eff}}^{(1/2)}$ we use both that $\mathcal{H}_{{\rm eff}}$ acts only inside a manifold and that $S$ has no matrix elements within a manifold. The general formulas are found e.g. in \cite{Hannes,Shavitt1980,Cohen1992}.\\
 The results for the effective Hamiltonian and the transformation matrix are given in the appendix \ref{Vleck}.
The non-vanishing matrix elements of the effective Hamiltonian, apart from the zeroth-order contributions in $g$, are
\begin{equation}
\left(\tilde{\mathcal{H}}_{\rm eff}\right)^{(1)}_{j{\rm e};(j+1){\rm g}}=-\frac{\hbar g\Delta_0}{\Delta_b}n_1(j)\equiv\hbar\Delta(j),
\end{equation}
and
\begin{eqnarray}
\left(\tilde{\mathcal{H}}_{\rm eff}\right)^{(2)}_{j{\rm e};j{\rm e}}&=&%
\hbar\left[W_1(j,\Omega)-W_{0}(j,\Omega)\right],
\end{eqnarray}
\begin{eqnarray}
\left(\tilde{\mathcal{H}}_{\rm eff}\right)^{(2)}_{j{\rm g};j{\rm g}}&=&%
\hbar\left[W_1(j,\Omega)+W_{0}(j+1,\Omega)\right].
\end{eqnarray}
We used as abbreviation
\begin{eqnarray}
n_1(j)&=&
\sqrt{\text{j}+1} \left(1+\frac{\sqrt{j}a_{-2}^{(j+1)}}{\sqrt{j+1}}+\frac{a_2^{(j)}
   \sqrt{j+2}}{\sqrt{j+1}}\right) \nonumber\\
&=&\sqrt{j+1}\left[1-\frac{3\alpha}{2\hbar\Omega}(j+1)\right]+\mathcal{O}(\alpha^2),
   \end{eqnarray}
and
\begin{equation}
W_1(j,\Omega)=-\frac{g^2 \varepsilon ^2 }{\Delta_b ^2 \Omega}+\frac{6\alpha g^2(2j+1)\varepsilon^2}{\hbar\Delta_b^2\Omega^2}+\mathcal{O}\left(\alpha ^2\right),
\end{equation}
\begin{equation}
W_{0}(j,\Omega)=-\frac{g^2\Delta_0^2j}{\Delta_b^2(\Delta_b+\Omega)}\left[1-\frac{3\alpha j (\Delta_b+2\Omega)}{\hbar\Omega(\Delta_b+\Omega)}\right]+\mathcal{O}\left(\alpha ^2\right).
\end{equation}
Therefore the effective Hamiltonian acquires in first order in the nonlinearity and in second order in the coupling the form:
\begin{widetext}
\begin{small}
\begin{eqnarray} \label{HeffMatrix}
&& \tilde{ \mathcal{ H}}_{\rm eff} = \hbar\cdot\\
&& \left(\begin{array}{ c|cc|c}
                                   			    \ddots &  &  & \\ \hline
								  & \frac{\Delta_b}{2}+j \Omega+\frac{3}{2\hbar}\alpha j(j+1) + W_1(j,\Omega) - W_{0}(j,\Omega) & \Delta(j) &  \\
								& & &\\
								 & \Delta(j) & -\frac{\Delta_b}{2} +(j+1) \Omega+\frac{3}{2\hbar}\alpha(j+1)(j+2) + W_1(j+1,\Omega) + W_{0}(j+2,\Omega) & \\
 & & & \\\hline
								 & & & \ddots
                                                        \end{array} \right)\nonumber
\end{eqnarray}
\end{small}
\end{widetext}
for the states $|j{\rm e}\rangle$ and $|(j+1) {\rm g}\rangle$. 
The ground state $|0\rangle_{\rm eff}\equiv|0{\rm g}\rangle$ is an eigenstate of $\tilde{\mathcal{H}}_{\rm eff}$ with eigenenergy: 
\begin{eqnarray}
E_0&=&\hbar(-\Delta_b/2 + W_1(0,\Omega) + W_{0}(1,\Omega)). %
\end{eqnarray}
Due to the doublet structure the blocks of the effective Hamiltonian are $2\times 2$ matrices and the corresponding eigenvectors are for $j\geq 0$:
\begin{eqnarray}\label{gl2}
\ket{2j+1}_{\rm eff}&=&\cos\left(\frac{\eta_j}{2}\right)\ket{(j+1)g}+\sin\left(\frac{\eta_j}{2}\right)\ket{j {\rm e}},\\
\ket{2j+2}_{\rm eff}&=&-\sin\left(\frac{\eta_j}{2}\right)\ket{(j+1)g}+\cos\left(\frac{\eta_j}{2}\right)\ket{j{\rm e}}\nonumber,
\end{eqnarray}
where $\tan\eta_j=\frac{2|\Delta(j)|}{\delta_j}$ and $0\leq\eta_j<\pi$. Moreover,
\begin{eqnarray}
\delta_{j}&=&\Delta_b-\Omega-\frac{3\alpha(j+1)}{\hbar}+W_1(j,\Omega)-W_1(j+1,\Omega)\nonumber\\&&-W_{0}(j,\Omega)-W_{0}(j+2,\Omega).
\end{eqnarray}
In turn the eigenstates of the qubit-nonlinear oscillator system are obtained from the transformation 
\begin{eqnarray}\label{gl8}
\ket{n}=\exp(-iS)\ket{n}_{\rm eff}.
\end{eqnarray}
Finally, the eigenenergies are then
\begin{eqnarray}\label{gl3}
E_{2j+1/2j+2}&=&\hbar(j+\frac{1}{2})\Omega+\frac{3}{2}\alpha(j+1)^2\nonumber\\&&
+\hbar( W_1(j,\Omega)+W_1(j+1,\Omega))/2\nonumber\\&&-\hbar W_{0}(j,\Omega)/2
+\hbar W_{0}(j+2,\Omega)/2\nonumber\\
&&\mp\frac{\hbar}{2}\sqrt{\delta_j^2+4|\Delta(j)|^2}.%
\end{eqnarray}
These eigenergies are also eigenenergies of $\tilde{\mathcal{H}}_{\rm TLS-NO}$ by construction and are depicted in figure \ref{EnSpec} (solid lines) for the case of an unbiased TLS, $\varepsilon=0$.
At finite coupling the degeneracy is lifted and we observe avoided crossings (solid lines in figure \ref{EnSpec}). Due to the coupling the resonance condition acquires a shift compared to (\ref{rescond}), the so-called Bloch-Siegert shift \cite{BlochSiegert},
\begin{eqnarray}\label{BSS}
\Omega&=&\Delta_b-\frac{3}{\hbar}\alpha(j+1)+W_1(j,\Delta_b)-W_1(j+1,\Delta_b)\nonumber\\&&-W_0(j,\Delta_b)-W_0(j+2,\Delta_b)\nonumber\\&&+3\frac{\alpha g^2\Delta_0^2}{2\hbar\Delta_b^4}(j+1)^2+\mathcal{O}(\alpha^2,g^4).
\end{eqnarray}
The resonance corresponds to $\delta_j=0$. We notice that the effect of the nonlinearity onto the Bloch-Siegert shift is very weak, namely at least of order $\mathcal{O}(\alpha g^2)$ and negligible for the values of nonlinearity and coupling we considered in the following.\\
At resonance, equation (\ref{BSS}), the minimal splitting of the former degenerate gap is:
\begin{eqnarray}\label{gl17}
E_{2j+2}-E_{2j+1}&=&\hbar\sqrt{j+1}g\frac{\Delta_0}{\Delta_b}\left[2-\frac{3}{\hbar\Omega}\alpha(j+1)\right]\nonumber\\&&+\mathcal{O}(\alpha^2,g^3).
\end{eqnarray}
We notice that at any point of our calculation we can set the nonlinearity to zero and reproduce the results obtained for the TLS-linear oscillator system \cite{Hannes}.%

\subsection{Dynamics of the qubit for the non-dissipative case}\label{Hamiltonian}
The time evolution of the qubit-nonlinear-oscillator system without bath is given by $\rho(t)=\exp(-\frac{i}{\hbar}\tilde{\mathcal{H}}_{\rm TLS-NO})\rho(0)\exp(+\frac{i}{\hbar}\tilde{\mathcal{H}}_{\rm TLS-NO})$ and therefore 
\begin{eqnarray}
\rho_{nm}(t)&=&\bra{n}\rho(t)\ket{m}=\exp(-i\omega_{nm}t)\rho_{nm}(0),
\end{eqnarray}
where $\omega_{nm}=\frac{1}{\hbar}\left(E_n-E_m\right)$. Consequently we obtain for the population difference in (\ref{PtQubitHO})
\begin{eqnarray}\label{gl1}
P(t)&=&p_0+\sum_{\stackrel{n,m}{n>m}}p_{nm}(0)\cos\omega_{nm}t,
\end{eqnarray}
where we introduced $p_0\equiv\sum_n p_{nn}(0)$. We observe from (\ref{gl1}) that the dynamics of the TLS is determined by an infinite number of oscillation frequencies rather than showing a single Rabi oscillation. %
To set the initial conditions we assume that the qubit starts in the state $\ket{R}$ and that the occupation numbers of the NO are Boltzmann distributed:
\begin{eqnarray}
\rho(0)&=&\ket{R}\bra{R}\frac{1}{Z_{{\rm NO}}}\exp(-\beta \mathcal{H}_{\rm NO}),
\end{eqnarray}
where
\begin{equation}
Z_{\rm NO}=\sum_{j=0}^\infty\exp[-\beta(\hbar\Omega j+\frac{3}{2}\alpha j(j+1))] 
\end{equation}
is the partition function of the oscillator and $\beta=(k_BT)^{-1}$ is the inverse temperature. In the TLS-NO eigenbasis we get:
\begin{eqnarray}\label{gl4}
\rho_{nm}(0)&=&\bra{n}\rho(0)\ket{m}\\
&=&\frac{1}{Z_{\rm NO}}\sum_{j=0}^\infty\exp[-\beta(\hbar\Omega j+\frac{3}{2}\alpha j(j+1))]\nonumber\\&&
\left[\cos\left(\frac{\Theta}{2}\right)\braket{n}{j {\rm g}}+\sin\left(\frac{\Theta}{2}\right)\braket{n}{ j \rm e}\right]\nonumber\\&&
\left[\cos\left(\frac{\Theta}{2}\right)\braket{j {\rm g}}{m}+\sin\left(\frac{\Theta}{2}\right)\braket{ j {\rm e}}{m}\right]\nonumber.
\end{eqnarray}
\subsubsection{Low temperature approximation} \label{LTAOB}
\begin{figure}[h]
\begin{flushright}
  \resizebox{\linewidth}{!}{
\includegraphics{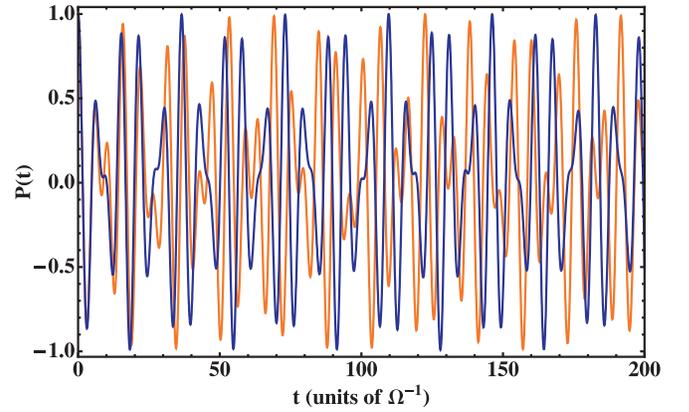}}
\end{flushright}
\begin{flushright}
  \resizebox{\linewidth}{!}{
\includegraphics{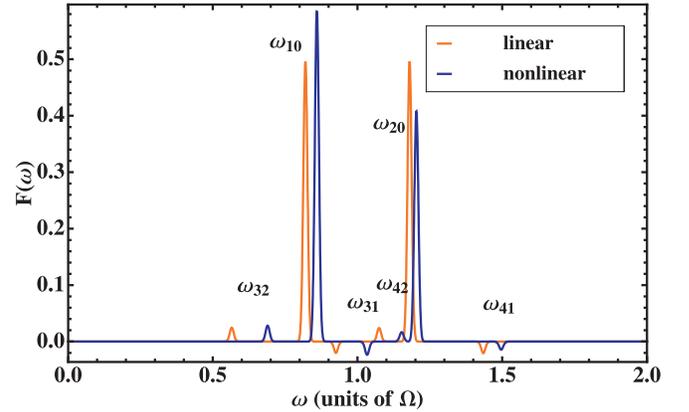}}
\end{flushright}
\caption{(Color online) Top: Dynamics of the population difference $P(t)$ for the unbiased, $\varepsilon=0$, qubit-nonlinear oscillator system at linear resonance $(\Omega=\Delta_0)$ (blue (dark gray) line). We choose a nonlinearity $\alpha=0.02\hbar\Omega$, a TLS-NO coupling $g=0.18\Omega$, and inverse temperature $\beta=10(\hbar\Omega)^{-1}$. For comparison we plotted the corresponding linear case (orange (light gray) line). Bottom: Fourier transform $F(\omega)$ of $P(t)$ for the unbiased system. The dominating frequencies are $\omega_{10}$ and $\omega_{20}$. To visualize the delta-functions, finite widths have artificially been introduced. \label{ft6}  }
\end{figure}

Equation (\ref{gl1}) allows us to describe the non-dissipative dynamics in terms of the approximate eigenenergies and eigenstates (\ref{gl8}) and (\ref{gl3}), which involve in this way all nonlinear oscillator states. Therefore the Hilbert space under consideration is infinite. To calculate $p_{nm}(0)$ and $p_{nn}(0)$ we need to know the structure of a matrix element such as $\braket{j,\{{\rm g/ e}\}}{n}=\bra{j,\{{\rm g/ e}\}}\exp(-iS)\ket{n}_{\rm eff}$. The $\ket{n}_{\rm eff}$ are themselves linear combinations of the uncoupled states $\ket{j,\{{\rm g/e}\}}$, see (\ref{gl2}).
 Because we calculated $\exp(-iS)$ up to second order in the coupling Hamiltonian $\mathcal{H}_{\rm Int}$, we find that the oscillator index $j$ can at most change by four, see appendix \ref{Vleck}. For typical experiments on qubits the temperature is restricted to the regime of $\beta^{-1}\ll\hbar\Omega,\hbar\Delta_b$. Due to the exponential function in (\ref{gl4}) high levels of the NO are only weakly populated and consequently we can truncate the infinite sum in equation (\ref{gl4}) for the matrix elements of the density matrix at initial time to $j=1$. This means that the lowest 12 $\{\ket{n}\}$ states enter (\ref{gl4}).\\
After a close analysis we observe, by inserting (\ref{gl4}) into (\ref{gl6}), that the coefficients $p_{nm}(0)$ with $n\geq 7$ are of higher than second order in $g$. The same is valid for $p_{50},p_{60},p_{55}$ and $p_{66}$. Thus those terms do not occur in the calculation of $P(t)$.
Of the remaining contributions we observe that those with $n=5,6$ are either at least of order $g\exp[-\beta(\hbar\Omega+3\alpha)]$ or of order $g^2\exp[-\beta(\hbar\Omega+3\alpha)]$ or of order $\alpha g^2$.  
Thus we can also disregard contributions from $p_{nm}$ for $n\geq 5$ for the parameters chosen in the following, i.e., in the considered low temperature regime it is enough to restrict to the five lowest eigenstates of $\tilde{\mathcal{H}}_{\rm TLS-NO}$. Therefore the number of possible oscillation frequencies $\omega_{nm}$ is reduced to 10, where $n,m=0,1,\dots,4$ and $n>m$.\\
In the following we show the dynamics of an unbiased TLS ($\varepsilon=0$), which results in vanishing of $p_0$, $p_{30}(0)$, $p_{40}(0)$, $p_{21}(0)$ and $p_{43}(0)$. Therefore we obtain:
\begin{eqnarray}
P(t)&=&p_{10}\cos(\omega_{10}t)+p_{20}\cos(\omega_{20}t)\\&&+p_{31}\cos(\omega_{31}t)+p_{41}\cos(\omega_{41}t)\nonumber\\
&&+p_{32}\cos(\omega_{32}t)+p_{42}\cos(\omega_{42}t)\nonumber.
\end{eqnarray}
Exemplarily we consider in the following the resonant case for the corresponding linear oscillator, where $\Omega=\Delta_b=\Delta_0$. This corresponds to a slightly detuned nonlinear-oscillator system. The resulting transition frequencies using (\ref{gl3}) are:
\begin{eqnarray}\label{gl16a}
\omega_{10}&=&\Omega-g+\frac{3 \alpha }{2 \hbar}+\frac{9 \alpha  g}{4 \hbar \Omega }+\frac{9 \alpha  g^2}{4 \hbar \Omega ^2},\\
\omega_{20}&=&\Omega+g+\frac{3 \alpha }{2 \hbar}-\frac{9 \alpha  g}{4 \hbar \Omega }+\frac{9 \alpha  g^2}{4 \hbar \Omega ^2}\nonumber,\\
\omega_{31}&=&\Omega+g(1-\sqrt{2})+\frac{9 \alpha }{2 \hbar}+\frac{9 \alpha  g}{4 \hbar \Omega }\left[2\sqrt{2}-1\right]+\frac{9 \alpha  g^2}{2 \hbar \Omega ^2}\nonumber,\\
\omega_{41}&=&\Omega+g(1+\sqrt{2})+\frac{9 \alpha }{2 \hbar}-\frac{9 \alpha  g}{4 \hbar \Omega }\left[2\sqrt{2}+1\right]+\frac{9 \alpha  g^2}{2 \hbar \Omega ^2}\nonumber,\\
\omega_{32}&=&\Omega-g(1+\sqrt{2})+\frac{9 \alpha }{2 \hbar}+\frac{9 \alpha  g}{4 \hbar \Omega }\left[2\sqrt{2}+1\right]+\frac{9 \alpha  g^2}{2 \hbar \Omega ^2}\nonumber,\\
\omega_{42}&=&\Omega-g(1-\sqrt{2})+\frac{9 \alpha }{2 \hbar}
-\frac{9 \alpha  g}{4 \hbar \Omega }\left[2\sqrt{2}-1\right]+\frac{9 \alpha  g^2}{2 \hbar \Omega ^2}\nonumber.
\end{eqnarray}
Due to the nonlinearity the six different oscillation frequencies in equation (\ref{gl16a}) are shifted to higher frequencies compared to the linear oscillator case $\alpha=0$. In contrast to the linear case they are no longer located symmetrically around $\Omega=\Delta_0$. The reason for this lies in the non-equidistant energy levels of the nonlinear oscillator alone and in the interplay of coupling and nonlinearity. %
The population difference $P(t)$ and its Fourier transform are shown in figure \ref{ft6}. As in the linear case, %
the dominating frequencies are $\omega_{10}$ and $\omega_{20}$. %
These correspond to transitions between the first and the second state of the qubit-NO-system and the ground state. In the linear oscillator case the weight of their peaks is almost equal, whereas with weak nonlinearities the peak corresponding to $\omega_{10}$ is more pronounced. This is due to the fact that the frequency corresponding to the more pronounced peak fits more accurately the resonance condition, which includes the 
Bloch-Siegert shift in (\ref{BSS}). The weight of the peaks can additionally be influenced by allowing a finite bias of the qubit, $\varepsilon\neq0$. The zero bias case was chosen here for simplicity.\\
From these graphs and equations (\ref{gl17}) and (\ref{gl16a}) we can read off first that the vacuum Rabi splitting is decreased for finite nonlinearity and second that the overall frequency shifts compared to the linear case are larger the higher the oscillator levels are involved if the coupling $g$ is not too large to overcome the effects caused by the nonlinearity.\\
\section{Influence of the environment}\label{env}
The knowledge about decoherence and dissipation processes entering in the qubit dynamics is essential for quantum computation. Therefore we consider now the qubit-nonlinear-oscillator system to be coupled to an environment and treat the full Hamiltonian $\mathcal{H}$.
\subsection{Master equation for the qubit-NO system}
As shown in section \ref{popdiff}, equation (\ref{PtQubitHO}), we need for the calculation of $P(t)$ the density matrix $\rho(t)$ of the qubit-nonlinear oscillator system. To take into account the effect of the bath we start from the Liouville equation for the full density matrix $W(t)$ of $\mathcal{H}$,
\begin{eqnarray}
i\hbar\frac{\partial W_{\rm I}(t)}{\partial t}&=&\left[\mathcal{H}_{\rm NO-B,{\rm I}}(t),W_{\rm I}(t)\right],
\end{eqnarray}
where the index $I$ denotes the interaction picture. Following \cite{Blum, Louisell} we arrive at a Born-Markov master equation for $\rho(t)$ being in the Schr\"odinger picture and expressed in the basis of the eigenstates of $\tilde{\mathcal{H}}_{\rm Q-NO}$:
\begin{eqnarray}\label{gl13}
\dot{\rho}_{nm}(t)&=&-i\omega_{nm}\rho_{nm}(t)+\pi\sum_{k,l}\mathcal{L}_{nm,kl}\rho_{kl}(t).
\end{eqnarray}
The first term includes the free dynamics, whereas the second accounts for the dissipative one. The Bloch-Redfield tensors are defined by:
\begin{eqnarray}\label{gl7}
\mathcal{L}_{nm,kl}&=&\left[G(\omega_{nk})N_{nk}-G(\omega_{lm})N_{ml}\right]y_{nk}y_{lm}\\&&
-\delta_{ml}\sum_{l'}G(\omega_{l'k})N_{l'k}y_{nl'}y_{l'k}\nonumber\\
&&+\delta_{nk}\sum_{k'}G(\omega_{lk'})N_{k'l}y_{lk'}y_{k'm}\nonumber,
\end{eqnarray}
with $N_{nm}=\frac{1}{2}\left[\coth(\hbar\beta\omega_{nm}/2)-1\right]$ and $y_{nm}=\bra{n}(B+B^\dagger)\ket{m}$. In the following we assume to have an Ohmic bath described by the spectral density $G(\omega)\equiv G_{\rm Ohm}(\omega)=\kappa\omega$.\\
For the derivation of the master equation besides the Born-Markov approximation more assumptions have been made. We only mention them briefly: first, we assume that the system and bath are initially uncorrelated (at $t=0$), i.e., $W(0)=\rho_{\rm I}(0)\rho_{\rm B}(0)$, where $\rho_{\rm B}(0)=Z_{\rm B}^{-1}\exp(-\beta \mathcal{H}_{\rm B})$
and $Z_{\rm B}$ is the partition function of the bath. Because the bath consists of infinite degrees of freedom we assume the effects of the interaction with the TLS-NO system on the bath to dissipate away quickly, such that the bath remains in thermal equilibrium for all times $t$:
$W_{\rm I}(t)=\rho_{\rm I}(t)\rho_B(0)$. Additionally an initial slip term is neglected, which occurs due to the sudden coupling of the system to the bath \cite{Weiss}. Finally we disregarded the Lamb-shift of the oscillation frequencies $\omega_{nm}$.
\subsection{Matrix elements}\label{secME}
The Redfield tensors, equation (\ref{gl7}), depend on the matrix elements $y_{nm}$ of the NO position operator in the TLS-NO eigenbasis. Using equation (\ref{gl8}) we rewrite $y_{nm}$ in the form:
\begin{eqnarray}
y_{nm}&=&\bra{n}y\ket{m}= \ _{\rm eff}\bra{n}\exp(iS)y\exp(-iS)\ket{m}_{\rm eff}\nonumber\\
&\equiv&\ _{\rm eff}\bra{n}\tilde{y}\ket{m}_{\rm eff}.
\end{eqnarray}
The effective states are given in (\ref{gl2}) as linear combinations of states of the $\left\{\ket{j{\rm g}};\ket{j{\rm e}}\right\}$ basis. In the following we show the different building blocks for $y_{nm}$. 
We can distinguish between different situations. First there are matrix elements where neither the qubit nor the oscillator state is changed, namely:
\begin{eqnarray}\label{l1}
\bra{j{\rm g}}\tilde{y}\ket{j{\rm g}}&=&-2(L_{LO0}(g)+L_{NO0}(j,\alpha,g)),\\
\bra{j{\rm e}}\tilde{y}\ket{j{\rm e}}&=&+2(L_{LO0}(g)+L_{NO0}(j,\alpha,g))\nonumber,
\end{eqnarray}
where $L_{LO0}(g)=g\varepsilon/\Delta_b\Omega$ and $L_{NO0}(j,\alpha,g)=-6\alpha g\varepsilon(2j+1)/\hbar\Delta_b\Omega^2$. These matrix elements contain contributions independent of the oscillator occupation number $j$ for zeroth order in the nonlinearity $\alpha$ and acquire a level dependence in first order.\\
A transition within the qubit is described by
\begin{eqnarray}\label{l3}
\bra{j{\rm g}}\tilde{y}\ket{j{\rm e}}&=&L_{LO0+}(g)+L_{NO0+}(\alpha,g)(2j+1).
\end{eqnarray}
Here we introduced abbreviations, given in appendix \ref{appME}, to show the actual order of the matrix elements involved. The notation is as follows: indices $LO$ and $NO$ refer to the linear or nonlinear oscillator, respectively. An additional index number, $\Delta j$, indicates that the nonlinear oscillator state is changed by $\Delta j$ quanta. We have elements where zero, one, two or three quanta are emitted or absorbed by the oscillator. Moreover we introduce indices $+/-$ or ${\rm g/e}$ which correspond to the TLS transition ${\rm g}\rightarrow {\rm e}$ or to ${\rm e}\rightarrow {\rm g}$, respectively, or to the qubit not changing from $\rm g$ or $\rm e$ configuration.\\
For the case $\Delta j=1$:
\begin{eqnarray}\label{l2}
\bra{j{\rm g}}\tilde{y}\ket{(j+1){\rm g}}&=&\sqrt{j+1}\left[1+(j+1)L_{NO}(\alpha)+\right.\nonumber\\
&&\left.L_{LO1}(g^2)+L_{NO1{\rm g}}(j,\alpha,g^2)\right],\nonumber\\
\bra{j{\rm e}}\tilde{y}\ket{(j+1){\rm e}}&=&\sqrt{j+1}\left[1+(j+1)L_{NO}(\alpha)-\right.\nonumber\\
&&\left.L_{LO1}(g^2)+L_{NO1{\rm e}}(j,\alpha,g^2)\right],
\end{eqnarray}
\begin{eqnarray}\label{l4}
\bra{j{\rm g}}\tilde{y}\ket{(j+1){\rm e}}&=&\sqrt{j+1}\left[L_{LO1+}(g^2)+\right.\nonumber\\
&&\left.L_{NO1+}(\alpha,g^2)(j+1)\right]\nonumber,\\
\bra{j{\rm e}}\tilde{y}\ket{(j+1){\rm g}}&=&\sqrt{j+1}\left[L_{LO1-}(g^2)+\right.\nonumber\\
&&\left.L_{NO1-}(\alpha,g^2)(j+1)\right],
\end{eqnarray} 
describe processes where an oscillator quantum is absorbed. All the matrix elements in (\ref{l3}), (\ref{l2}) and in (\ref{l4}) contain both zeroth-order as well as first-order contributions in the nonlinearity. Additionally, due to the fact that the states of the NO are linear combinations of the linear oscillator states, see equation (\ref{gl9}), additional transitions involving a change of the oscillator state by more than one quantum are allowed. They correspond to $\Delta j=2$, $\Delta j=3$ and read as
\begin{eqnarray}\label{gl30}
\bra{j{\rm g}}\tilde{y}\ket{(j+2){\rm g}}&=&\sqrt{(j+1)(j+2)}L_{NO2}(\alpha,g),\\
\bra{j{\rm g}}\tilde{y}\ket{(j+2){\rm e}}&=&\sqrt{(j+1)(j+2)}L_{NO2+}(\alpha,g),\nonumber\\
\bra{j{\rm e}}\tilde{y}\ket{(j+2){\rm g}}&=&\sqrt{(j+1)(j+2)}L_{NO2-}(\alpha,g),\nonumber\\
\bra{j{\rm e}}\tilde{y}\ket{(j+2){\rm e}}&=&-\sqrt{(j+1)(j+2)}L_{NO2}(\alpha,g),\nonumber\\
\bra{j{\rm g}}\tilde{y}\ket{(j+3){\rm g}}&=&\sqrt{(j+1)(j+2)(j+3)}\left[L_{NO3}(\alpha,g^2)\right.\nonumber\\
&&\left.-L_{NO}(\alpha)/2\right],\nonumber\\
\bra{j{\rm g}}\tilde{y}\ket{(j+3){\rm e}}&=&\sqrt{(j+1)(j+2)(j+3)}L_{NO3+}(\alpha,g^2),\nonumber\\
\bra{j{\rm e}}\tilde{y}\ket{(j+3){\rm g}}&=&\sqrt{(j+1)(j+2)(j+3)}L_{NO3-}(\alpha,g^2),\nonumber\\
\bra{j{\rm e}}\tilde{y}\ket{(j+3){\rm e}}&=&\sqrt{(j+1)(j+2)(j+3)}\left[-L_{NO3}(\alpha,g^2)\right.\nonumber\\
&&\left.-L_{NO}(\alpha)/2\right]\nonumber.
\end{eqnarray}
Notice that all terms in (\ref{gl30}) vanish when $\alpha=0$. %
The terms in (\ref{l2}) and (\ref{gl30}) involving no change in the qubit and a change in the oscillator by $\Delta j=1$ and $\Delta j=3$ quanta contain $g$-independent nonlinear contributions. The interplay of nonlinearity and coupling in lowest order can be observed in $\bra{j{\rm{g}}}\tilde{y}\ket{j{\rm{e}}}$, and in the terms involving an oscillator level change by 2. Additionally at the degeneracy point, $\varepsilon=0$, $L_{LO0}(g)$, $L_{NO0}(j,\alpha,g)$, $L_{LO1\pm}(g^2)$, $L_{NO2}(\alpha,g)$, $L_{NO1\pm}(\alpha,g^2)$, $L_{NO3\pm}(\alpha,g^2)$, and parts of $L_{NO1\{g/e\}}(j,\alpha,g^2)$ vanish. We are now able to calculate the matrix elements $y_{nm}$. They are given in appendix \ref{appME}.
\subsection{Dissipative dynamics}
To calculate $P(t)$ we have to solve the system of coupled differential equations (\ref{gl13}). When several TLS-NO levels are involved an exact solution can only be found numerically. Hence, in the remaining of this section we discuss three different approximation schemes, two based on the full secular approximation (FSA) applied to (\ref{gl13}) and one based on a partial secular approximation (PSA). We then compare the so obtained analytical predictions with the exact numerical solution of (\ref{gl13}).
\subsubsection{Full secular approximation (FSA)}
 We define:
\begin{eqnarray}\label{gl11}
\rho_{nm}(t)&=&\exp(-i\omega_{nm}t)\sigma_{nm}(t),
\end{eqnarray}
which, inserted in (\ref{gl13}), enables us to obtain a set of differential equations for $\dot{\sigma}_{nm}(t)$:
\begin{equation}\label{gl10}
\dot{\sigma}_{nm}(t)=\pi\sum_{kl}\mathcal{L}_{nm,kl}\exp[i(\omega_{nm}-\omega_{kl})t]\sigma_{kl}(t).
\end{equation}
The FSA consists of neglecting fast rotating terms in equation (\ref{gl10}) such that only terms survive where $\omega_{nm}-\omega_{kl}$ vanishes. This allows an effective decoupling of diagonal and off-diagonal elements such that
\begin{subequations}
\begin{eqnarray}
\dot{\sigma}_{nn}(t)&=&\pi\sum_{k}\mathcal{L}_{nn,kk}\sigma_{kk}(t),\label{gl14a}\\
\dot{\sigma}_{nm}(t)&=&\pi\mathcal{L}_{nm,nm}\sigma_{nm}(t)\mbox{ for }n\neq m.\label{gl14b}
\end{eqnarray}
\end{subequations}
The off-diagonal elements are determined by
\begin{eqnarray}
\sigma_{nm}(t)&=&\sigma_{nm}^0\exp(\pi \mathcal{L}_{nm,nm}t),
\end{eqnarray}
which results with (\ref{gl11}) in
\begin{eqnarray}\label{gl12}
\rho_{nm}(t)&=&\rho_{nm}^0\exp(\pi \mathcal{L}_{nm,nm}t)\exp(-i\omega_{nm}t).
\end{eqnarray}
The separation of the oscillatory motion of the dynamics from the relaxation one allows us to divide (\ref{PtQubitHO}) into two parts
\begin{eqnarray}\label{gl31}
P(t)&=&P_{{\rm relax}}(t)+P_{{\rm dephas}}(t),
\end{eqnarray}
where $P_{{\rm relax}}(t)=\sum_n p_{nn}(t)$ is the relaxation contribution and $P_{{\rm dephas}}(t)=\sum_{n>m}p_{nm}(t)$ is the dephasing part. Inserting (\ref{gl12}) in the last expression and using (\ref{gl6}), we obtain:
\begin{equation}
P_{{\rm dephas}}(t)=\sum_{n>m}p_{nm}(0)\exp(-\Gamma_{nm}t)\cos(\omega_{nm}t),
\end{equation}
where the dephasing rates are determined by $\Gamma_{nm}\equiv-\pi\mathcal{L}_{nm,nm}$. The actual form of the dephasing coefficients $\mathcal{L}_{nm,nm}$ can be found in appendix \ref{appdeph} and the initial conditions $\rho^0_{nm}=\sigma^0_{nm}=\rho_{nm}(0)$ are defined in (\ref{gl4}). The diagonal elements are more difficult to obtain, since the coupled system of differential equations in (\ref{gl14a}) has to be solved. To proceed we restrict ourselves in this section again to the physical relevant low temperature case, such that the highest qubit-nonlinear oscillator state involved is the eigenstate $\ket{4}$.  Calculating the rate coefficients accompanied with the five lowest eigenstates, we observe that there are only eight independent ones due to the structure of the rate coefficients. These are $\mathcal{L}_{00,11}$, $\mathcal{L}_{00,22}$, $\mathcal{L}_{11,22}$, $\mathcal{L}_{11,33}$, $\mathcal{L}_{11,44}$, $\mathcal{L}_{22,33}$, $\mathcal{L}_{22,44}$, and $ \mathcal{L}_{33,44}$. In general they are determined by
\begin{eqnarray} \label{IndependentRateCoeff}
  \mathcal{L}_{jj,kk} = 2 G(\omega_{jk}) N_{jk} y_{jk}^2 \quad {\rm with} \quad j<k,
\end{eqnarray} 
where $j$ and $k$ adopt the above values. Furthermore, $\mathcal{L}_{00,33}$, $\mathcal{L}_{00,44}$, $\mathcal{L}_{33,00}$ and  $\mathcal{L}_{44,00}$ are disregarded, because they are at least of order $\mathcal{O}(g^4)$. The remaining rate coefficients are combinations of the above. We find that
\begin{eqnarray} \label{DependentRateCoeff}
  \mathcal{L}_{kk,jj} &=& \mathcal{L}_{jj,kk} + 2 G(\omega_{jk}) y_{jk}^2 \\&=& (N_{jk}+1) 2 G(\omega_{jk}) y_{jk}^2,\nonumber
\end{eqnarray} 
and

\begin{eqnarray} \label{gl16}
    \mathcal{L}_{00,00}& =& - \mathcal{L}_{11,00} -  \mathcal{L}_{22,00}, \label{DependentRates1}\\
    \mathcal{L}_{11,11} &=& - \mathcal{L}_{00,11} -  \mathcal{L}_{22,11} - \mathcal{L}_{33,11} - \mathcal{L}_{44,11},\nonumber \\
    \mathcal{L}_{22,22} &=& - \mathcal{L}_{00,22} -  \mathcal{L}_{11,22} - \mathcal{L}_{33,22} - \mathcal{L}_{44,22}, \nonumber\label{DependentRates2}\\
    \mathcal{L}_{33,33} &=& - \mathcal{L}_{11,33} -  \mathcal{L}_{22,33} - \mathcal{L}_{44,33}, \nonumber\\
    \mathcal{L}_{44,44} &=& - \mathcal{L}_{11,44} -  \mathcal{L}_{22,44} - \mathcal{L}_{33,44}.\nonumber \label{DependentRates3}
\end{eqnarray}
\subsubsection*{Low temperature approximation (LTA)}
Despite the above relations equation (\ref{gl14a}) is too complicated to be solved analytically. Therefore an additional approximation is applied:  we consider the factor $N_{nm}+1=\frac{1}{2}\left[\coth(\hbar\beta\omega_{nm}/2)+1\right]$ with $n<m$ in equation (\ref{DependentRateCoeff}) and use that ${\rm lim}_{x\rightarrow -\infty}\coth(x/2)=-1$ is reached exponentially fast.\\ %
The terms containing this factor are neglected in the following. As we consider only the lowest five levels, this amounts to require ${\rm max}\{\omega_{nm}\}=|\omega_{14}|\gg k_b T$. Using equation (\ref{gl17}) we observe that $\omega_{12}\propto g$ and $\omega_{34}\propto g$. For this reason and due to the structure of $y_{nm}$ given in equation (\ref{ynm1}) the rates $\mathcal{L}_{11,22}$ and $\mathcal{L}_{33,44}$ are at least of order $\mathcal{O}(g^3)$ and can be neglected. With equation (\ref{gl16}) the rate matrix $\mathcal{L}_{\rm relax}$ associated to (\ref{gl14a}) becomes
\begin{widetext}\begin{eqnarray} \label{SimplifiedMasterEq}
 \mathcal{L}_{\rm relax} =
  && \left( \begin{array}{ccccc}
                  0 & \mathcal{L}_{00,11} & \mathcal{L}_{00,22} & 0 & 0 \\
                  0 & -\mathcal{L}_{00,11} & 0 & \mathcal{L}_{11,33} & \mathcal{L}_{11,44} \\
                  0 & 0 & -\mathcal{L}_{00,22} & \mathcal{L}_{22,33} & \mathcal{L}_{22,44}\\
                  0 & 0 & 0 & - \mathcal{L}_{11,33} - \mathcal{L}_{22,33} & 0  \\
                  0 & 0 & 0 & 0 & - \mathcal{L}_{11,44} - \mathcal{L}_{22,44}
                \end{array}
       \right).\nonumber
\end{eqnarray} \end{widetext}
The eigenvalues and eigenvectors of this matrix and the associated time evolution of the elements $\sigma_{nn}(t)$ are given in appendix \ref{AppDiagonalElements}. In contrast to the simple analytic expression for the dephasing part the relaxation rate is not easy to extract as $P_{\rm relax}(t)=\sum_n p_{nn}(t)$ consists of a sum of several exponential functions, cf. (\ref{gl6}) and appendix \ref{AppDiagonalElements}. However an analytical formula for $P(t)$ can be provided using (\ref{gl31}).

\subsubsection*{Smallest eigenvalue approximation (SEA)}
\begin{figure}[h]
\begin{flushright}
  \resizebox{\linewidth}{!}{
  \includegraphics{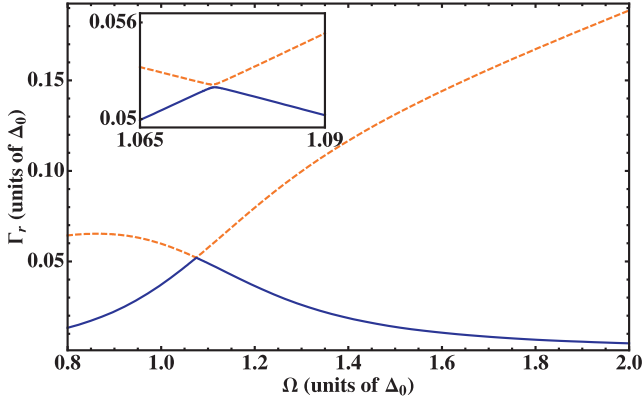}
\begin{large}                                                                                  \end{large}}
\end{flushright}
\caption{(Color online) The relaxation rate $\Gamma_r$ given in equation (\ref{RelaxationRate}) drawn against the oscillator fequency $\Omega$ (continuous blue (dark gray) line). We used $\epsilon=0.5\Delta_0$, corresponding to a frequency splitting $\Delta_b=1.118\Delta_0$, coupling $g=0.18\Delta_0$ and the nonlinearity $\alpha=0.02\hbar\Delta_0$. The damping constant is $\kappa=0.0154$ and $\beta=10(\hbar\Delta_0)^{-1}$. At resonance ($\Omega=\Delta_b-3\alpha/\hbar+\frac{g^2\Delta_0^2}{\Delta_b^3}$) $\Gamma_r$ is maximal. For comparison also the second lowest eigenvalue is plotted (orange (light gray) dashed line). The inset shows the two eigenvalues close to resonance.\label{Rate}}
\end{figure}

In order to get a better insight into the effect of the relaxation mechanism, we consider the long-time dynamics of the system. This means that the we direct our attention to the smallest eigenvalue of the relaxation coefficients, which dominates at long time, rather than to tackle the many relaxation contributions involved in the populations $\sigma_{nn}(t)$. We do not make the low temperature approximation discussed above. We restrict for simplicity to the three lowest qubit-NO eigenstates $\ket{0}$, $\ket{1}$, $\ket{2}$ in (\ref{gl14a}) and obtain using (\ref{gl16}):
\begin{small}
\begin{eqnarray} \label{RelaxationMatrix}
 && \mathcal{L}_{\rm relax} =\\
&&  \left( \begin{array}{ccc}
                  - \mathcal{L}_{11,00} - \mathcal{L}_{22,00} & \mathcal{L}_{00,11} & \mathcal{L}_{00,22}  \\
                  \mathcal{L}_{11,00} & -\mathcal{L}_{00,11} - \mathcal{L}_{22,11} &  \mathcal{L}_{11,22} \\
                  \mathcal{L}_{22,00} & \mathcal{L}_{22,11} & -\mathcal{L}_{00,22} -\mathcal{L}_{11,22}
                \end{array}
       \right).\nonumber
\end{eqnarray}
\end{small} 
We do not neglect $\mathcal{L}_{11,22}$ and $\mathcal{L}_{22,11}$, even if they are at least of order $\mathcal{O}(g^3)$, because these contributions lift the degeneracy of the two lowest eigenvalues at resonance (see figure \ref{Rate}). The smallest eigenvalue is:
\begin{eqnarray} \label{RelaxationRate}
    &&  \Gamma_{\rm r} \equiv\\
&& - \frac{\pi}{2} \biggl\{ - \sum_{n\neq m} \mathcal{L}_{nn,mm}  + \biggl[  \biggl( \sum_{n\neq m} \mathcal{L}_{nn,mm}  \biggr)^2\nonumber
\\&& -4 (\mathcal{L}_{00,11} \mathcal{L}_{00,22} + \mathcal{L}_{11,00} \mathcal{L}_{00,22}       + \mathcal{L}_{00,11} \mathcal{L}_{11,22} + \nonumber \\
&& \mathcal{L}_{11,00} \mathcal{L}_{11,22} + \mathcal{L}_{00,11} \mathcal{L}_{22,00} + \mathcal{L}_{11,22} \mathcal{L}_{22,00} \nonumber\\
&&     + \mathcal{L}_{22,11} \mathcal{L}_{00,22} + \mathcal{L}_{11,00} \mathcal{L}_{22,11} + \mathcal{L}_{22,00} \mathcal{L}_{22,11} ) \biggr]^{1/2} \biggr\}.\nonumber
\end{eqnarray}
Additional detuning allows for a further simplification: $\Gamma_r\approx\pi\mathcal{L}_{00,22}$ for $\Omega+3\alpha/\hbar-\frac{g^2\Delta_0^2}{\Delta_b^3}<\Delta_b$ and $\Gamma_r\approx\pi\mathcal{L}_{00,11}$ for $\Omega+3\alpha/\hbar-\frac{g^2\Delta_0^2}{\Delta_b^3}>\Delta_b$.
In figure \ref{Rate} the relaxation rate $\Gamma_r$ in (\ref{RelaxationRate}) is plotted as a function of the linear oscillator frequency $\Omega$. It is maximal at resonance, whereas it decays for $\Omega$ being detuned from resonance.
 Additionally we plotted the second smallest eigenvalue of (\ref{RelaxationMatrix}) for comparison (dashed orange (light gray) line in figure \ref{Rate}).\\
In the long-time limit it then holds:
\begin{equation}
  P_{{\rm relax}}(t) = (p_0 - p_\infty) e^{-\Gamma_{\rm r}t} + p_\infty, 
\end{equation} 
where, like in section \ref{Hamiltonian}, $p_0 \equiv \sum_n p_{nn}(0)$. To obtain $p_\infty$ we have in principle to find the steady-state solution of (\ref{gl14a}). Here, we just assume for $t \to \infty$ a Boltzmann distribution for the TLS-NO system, so that
  $\rho_{nn}(\infty) = Z_{\rm TLS-NO}^{-1} \exp(-\beta E_n)$ with $Z_{\rm TLS-NO} = \sum_{n} \exp(-\beta E_{n})$.
Thus,
\begin{eqnarray}
  p_\infty &=& \sum_n \sum_j \left\{ \cos \Theta \biggl[ \braket{j {\rm g}}{n}^2 - \braket{j {\rm e}}{n}^2 \biggr]
\right.\\&&\left.  + 2 \sin \Theta \braket{j {\rm g}}{n} \braket{j {\rm e}}{n} \right\} \rho_{n n}(\infty)\nonumber \label{pinf}.
\end{eqnarray}
The formula for the long-time dynamics is then obtained,
        \begin{eqnarray} \label{PSimpleFSA}
          P(t) &=& (p_0 - p_\infty) \exp(-\Gamma_{\rm r}t) + p_\infty +\\&& \sum_{n>m} p_{nm}(0) \exp(-\Gamma_{nm} t) \cos(\omega_{nm} t).\nonumber
        \end{eqnarray}
To get further insight on the dominant frequencies we evaluate the Fourier transform of (\ref{PSimpleFSA}) according to
\begin{equation}
  F(\omega) = 2 \int_0^\infty dt \cos \omega t P(t),
\end{equation} 
yielding
\begin{eqnarray} \label{FSimpleFSA}
 &&F(\omega) =\\&& 2 (p_0 - p_\infty) \frac{\Gamma_{\rm r}}{\omega^2 + \Gamma_{\rm r}^2} + 2 \pi p_\infty \delta(\omega) + \sum_{n<m} p_{nm}(0) \Gamma_{mn}\nonumber\\
  &&\times \left[ \frac{1}{\Gamma_{mn}^2 + (\omega_{mn} + \omega)^2} + \frac{1}{\Gamma_{mn}^2 + (\omega_{mn} - \omega)^2}   \right].\nonumber
\end{eqnarray}

\subsubsection{Partial secular approximation (PSA)} \label{PSA}
The PSA is an improvement to the FSA, where one accounts for corrections to the equations for the coherences due to dominant rotating terms $\omega_{nm} - \omega_{kl}$ in (\ref{gl10}). The equation for the populations is still given by (\ref{gl14a}). At low temperatures the dominant correction to the FSA comes from transitions involving the quasi-degenerate states $\ket{1}$ and $\ket{2}$.
To solve the off-diagonal part we have to determine $\sigma_{01}$, $\sigma_{02}$, $\sigma_{13}$, $\sigma_{23}$, $\sigma_{14}$, and $\sigma_{24}$. With  (\ref{gl10}) the system of equations is
\begin{equation}
  \dot \rho_{nm} (t) = (-  i \omega_{nm} + \pi \mathcal{L}_{nm,nm}) \rho_{nm}(t) + \pi \mathcal{L}_{nm,jk} \rho_{jk}(t),
\end{equation}
\begin{equation} 
  \dot \rho_{jk} (t) = \pi \mathcal{L}_{jk,nm} \rho_{nm}(t) + (- i \omega_{jk} + \pi \mathcal{L}_{jk,jk}) \rho_{jk}(t)
\end{equation}
with $\{ (nm), (jk) \}=\{(01);(02)\}$,$\{(13);(23)\}$, or $\{(14);(24)\}. $
The solution is
\begin{eqnarray} 
  \rho_{nm} &=& c_{nm,jk}^{(+)} v_{nm,jk}^{(+)} \exp(\lambda_{nm,jk}^{(+)} t)\\&& + c_{nm,jk}^{(-)} v_{nm,jk}^{(-)} \exp(\lambda_{nm,jk}^{(-)} t), \nonumber\\
  \rho_{jk} &=& c_{nm,jk}^{(+)} \exp(\lambda_{nm,jk}^{(+)} t) + c_{nm,jk}^{(-)} \exp(\lambda_{nm,jk}^{(-)} t), \nonumber
\end{eqnarray}
where the oscillation frequencies and the decay of the off-diagonal elements are given by \cite{Hannes}
\begin{widetext}
\begin{equation} \label{DephasingEigenvalues}
  \lambda_{nm,jk}^{(+/-)} = \frac{1}{2} \left[ \pi (\mathcal{L}_{nm,nm} + \mathcal{L}_{jk,jk}) -  i (\omega_{nm} + \omega_{jk}) \pm R_{nm,jk}   \right]
\end{equation} 
with
\begin{equation}
   R_{nm,jk} = \sqrt{\left[ \pi (\mathcal{L}_{nm,nm} - \mathcal{L}_{jk,jk}) -  i (\omega_{nm} - \omega_{jk}) \right]^2 + 4 \pi^2 \mathcal{L}_{nm,jk} \mathcal{L}_{jk,nm}}.
\end{equation} 
The amplitudes of the oscillations are given through the coefficients
\begin{equation}
   c_{nm,jk}^{(+/-)} = \pm \frac{2 \pi \mathcal{L}_{jk,nm} \rho_{nm}^0 - \rho_{jk}^0 \left[ \pi (\mathcal{L}_{nm,nm} - \mathcal{L}_{jk,jk}) -  i (\omega_{nm} - \omega_{jk}) \mp R_{nm,jk} \right]}{2 R_{nm,jk}}
\end{equation}
and
\begin{equation}
   v_{nm,jk}^{(+/-)} = \frac{1}{2 \pi\mathcal{L}_{jk,nm}} \left[ \pi (\mathcal{L}_{nm,nm} - \mathcal{L}_{jk,jk}) -  i (\omega_{nm} - \omega_{jk}) \pm R_{nm,jk} \right].
\end{equation}
\end{widetext}
We can calculate analytically the relaxation and dephasing part of $P(t)$. While the FSA allows a simple form for the dephasing rates, $\Gamma_{nm} = -\pi \mathcal{L}_{nm,nm}$, the PSA one is much more involved. As in case of the SEA the smallest eigenvalue dominates the dephasing behavior. The corresponding Bloch-Redfield tensors are found in appendix \ref{appdeph}. 

\section{Numerical versus analytical predictions}\label{secRes}
In the following we compare the results for the dynamical quantity $P(t)$ and its Fourier transform, obtained by a numerical solution of (\ref{gl13}), with the predictions of the approximations from section \ref{env}. %
\subsection{Low temperature}
\begin{figure}[h!]
\begin{flushright}
  \resizebox{\linewidth}{!}{
\includegraphics{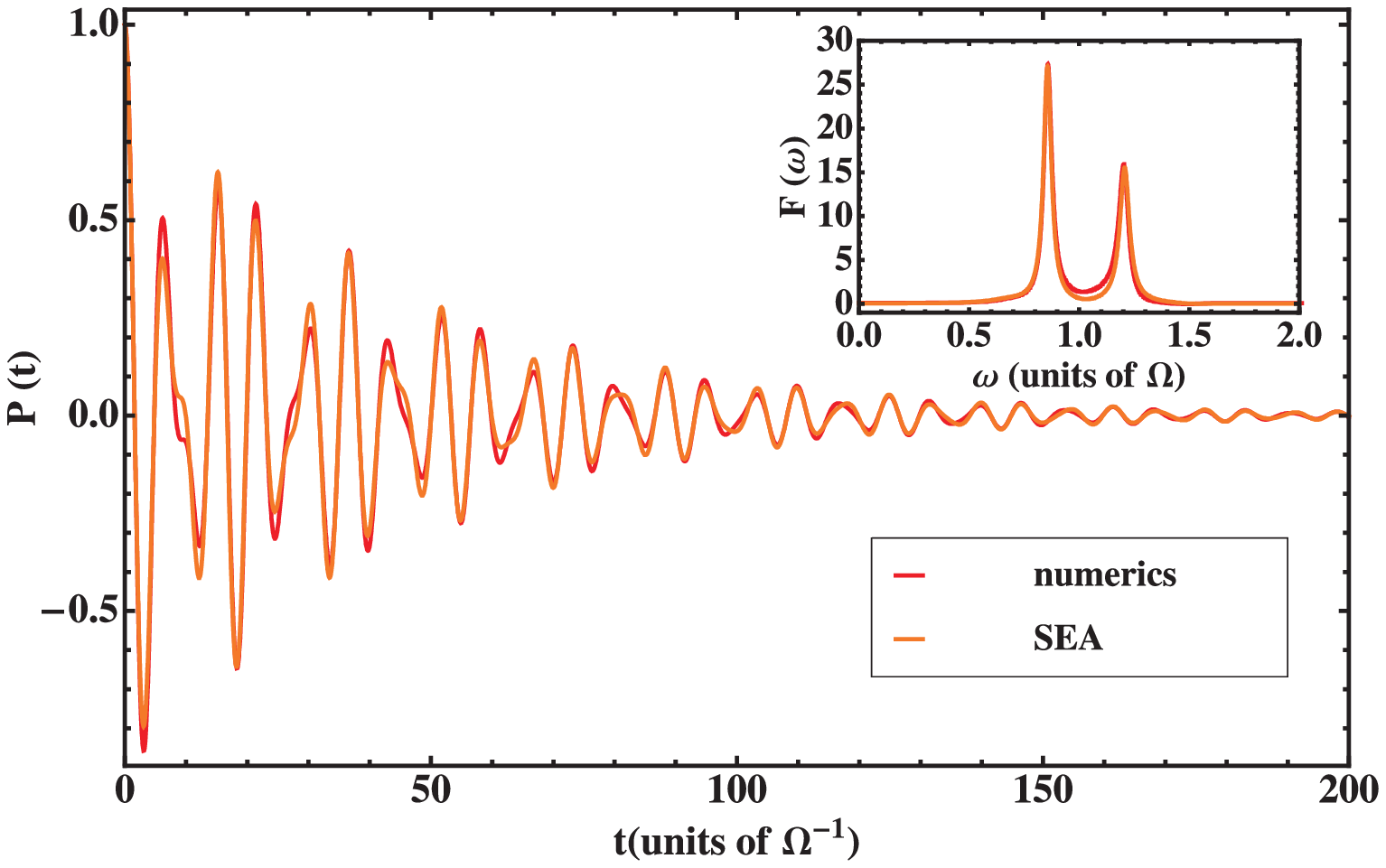}}
\end{flushright}
\begin{flushright}
  \resizebox{\linewidth}{!}{
  \includegraphics{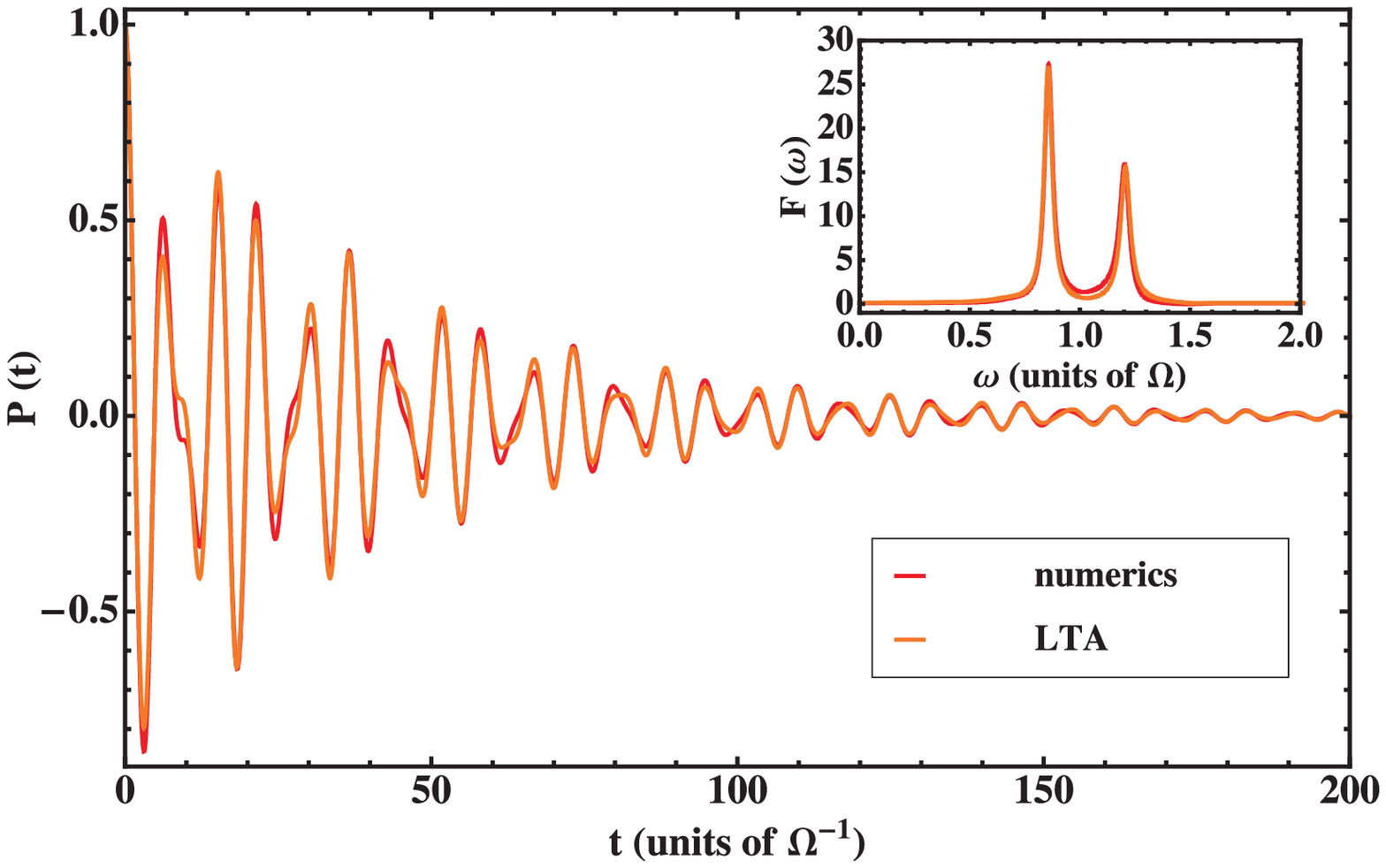}}
\end{flushright}
\begin{flushright}
  \resizebox{\linewidth}{!}{
  \includegraphics{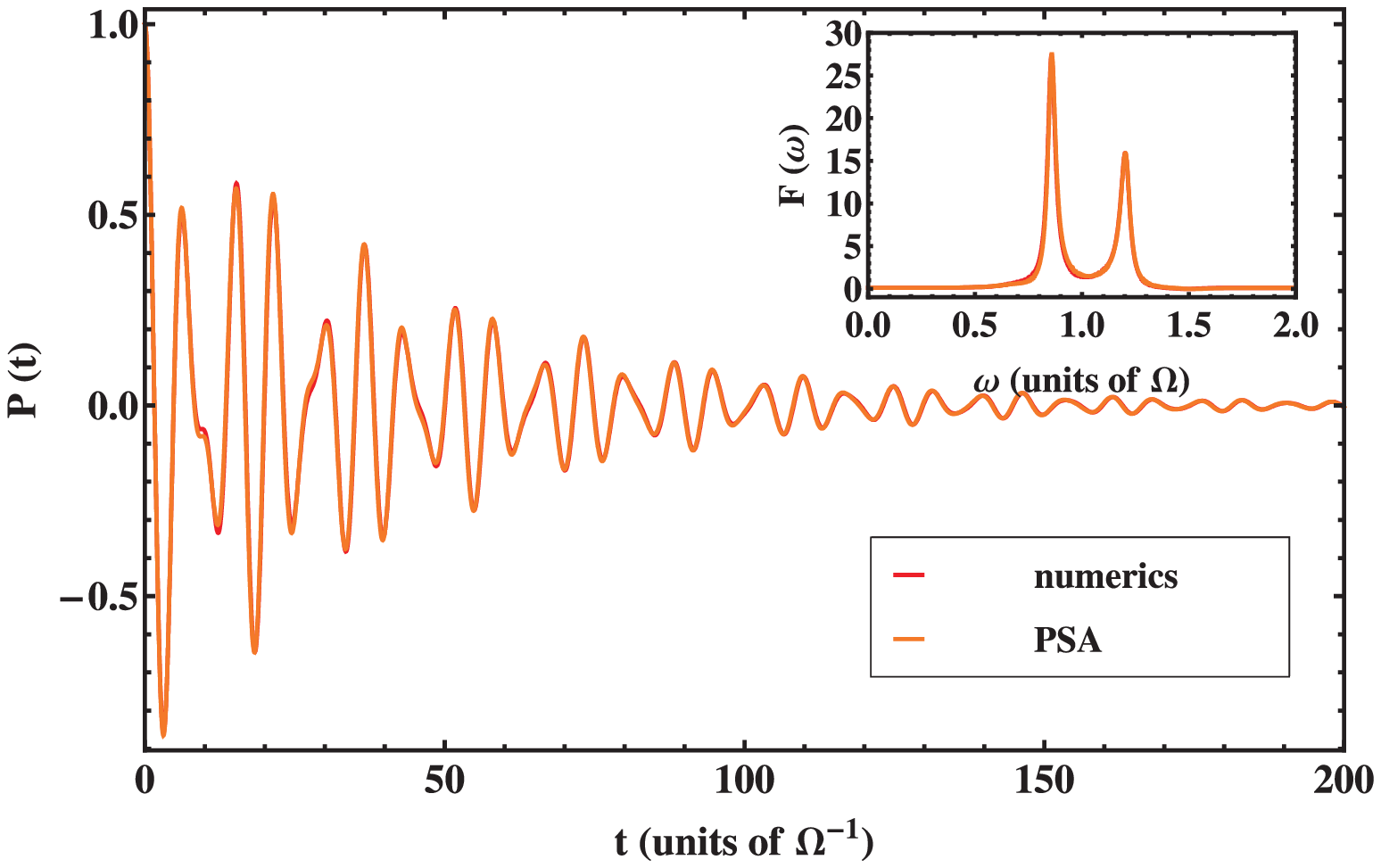}}
\end{flushright}
\caption{(Color online) Comparison of the behaviour of $P(t)$  and its Fourier transform $F(\omega)$ as obtained from the numerically exact solution (red (dark gray) line)  and the three approximation schemes (orange (light gray) line) discussed in the text. Top: Smallest eigenvalue approximation (SEA), Middle: Low temperature approximation (LTA) and Bottom: Partial secular approximation (PSA). The chosen parameters are: $\alpha=0.02\hbar\Omega$, $g=0.18\Omega$, $\varepsilon=0\Omega$, $\kappa=0.0154$ and $\beta=10(\hbar\Omega)^{-1}$. The dynamics is well reproduced within all approximations, however the agreement of the PSA with the exact numerics is the best. In the corresponding Fourier spectrum almost no deviations occur for all three approaches.\label{CompPexactPApprox2}  }
\end{figure}

\begin{figure}[h!]
\begin{flushright}
  \resizebox{\linewidth}{!}{
  \includegraphics{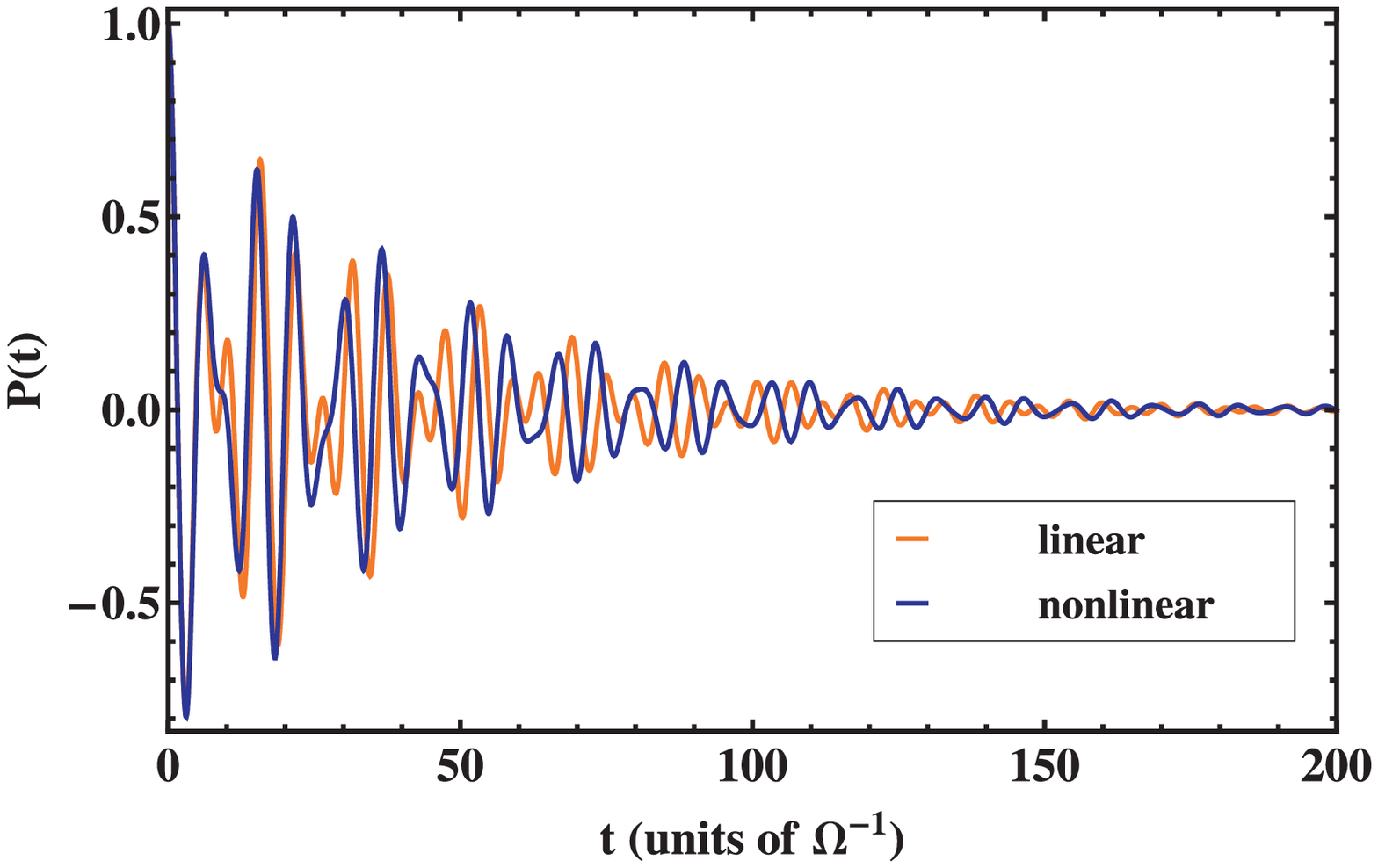}}
\end{flushright}
\begin{flushright}
  \resizebox{\linewidth}{!}{
  \includegraphics{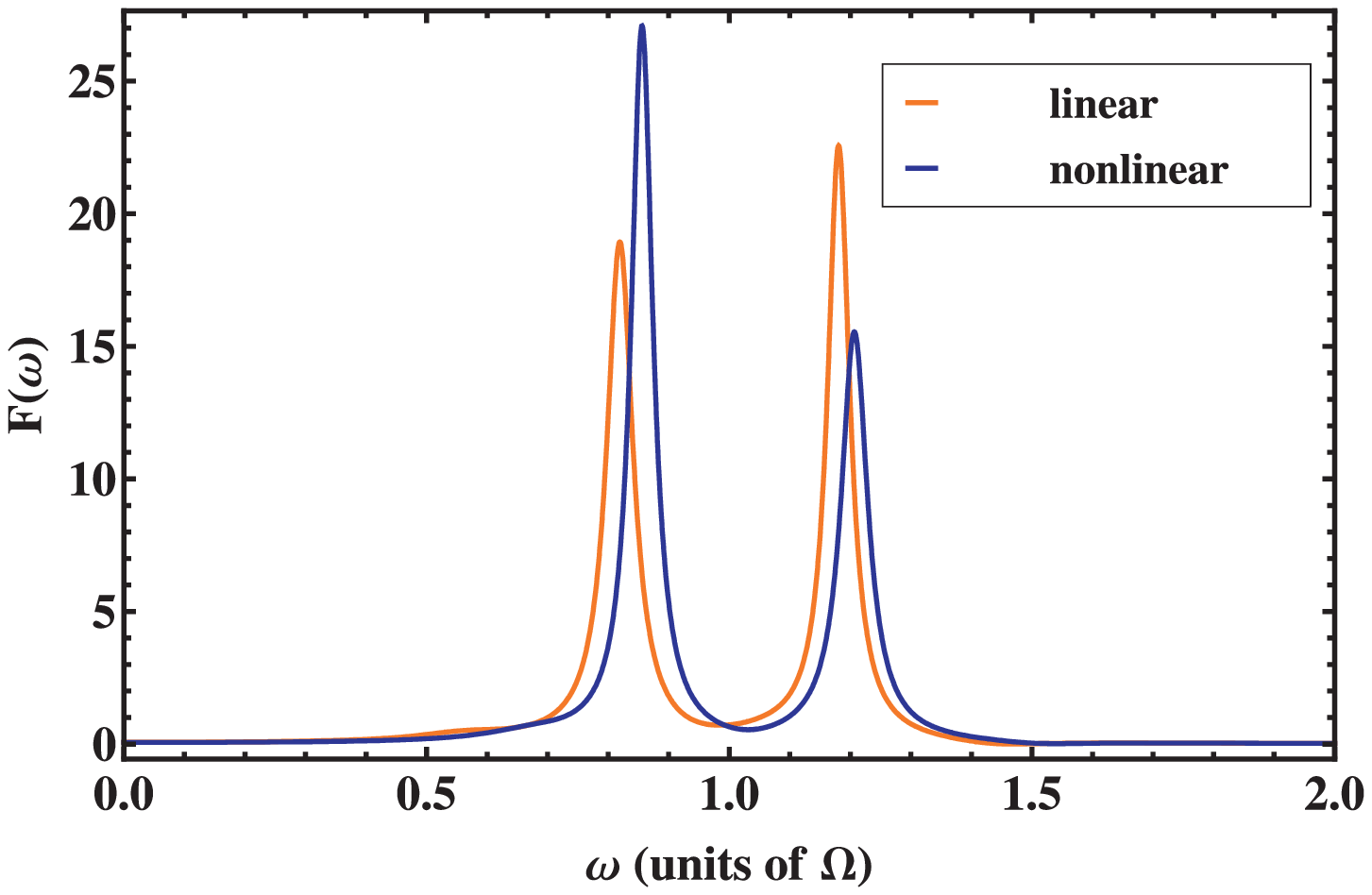}}
\end{flushright}
\caption{(Color online) Top: $P(t)$ for the linear (orange (light gray) line) and nonlinear (blue (dark gray) line) case using the parameters: $\alpha=0$ or $\alpha=0.02\hbar\Omega$, respectively, and  $g=0.18\Omega$, $\kappa=0.0154$, $\varepsilon=0$, $\Delta_b=\Omega$, $\beta=10(\hbar\Omega)^{-1}$. Bottom: Corresponding Fourier transform $F(\omega)$.\label{FFig1}  }
\end{figure}

We start by focusing on low temperatures $\beta=10/(\hbar\Omega)$ and compare the results for all three approaches (SEA, LTA, and PSA) to the numerical solution in figure \ref{CompPexactPApprox2}. We recognize that the dynamics and the corresponding Fourier spectrum are well reproduced within the simple SEA approach as well as in the two LTA and PSA treatments and determined by the superposition of two oscillations. The best approximation is the PSA. In the following we use the SEA approach due to its simpler analytic form.\\
To determine the effects of the nonlinearity onto the qubit dynamics we compare $P(t)$ and $F(\omega)$ with the corresponding linear case in figure \ref{FFig1}. We choose $\Omega=\Delta_b$.
Both in the nonlinear and in the corresponding linear case two oscillation frequencies are dominant. Due to the Bloch-Siegert shift, see (\ref{BSS}), in both cases $\Omega=\Delta_b$ is not the exact resonance condition. However, in the nonlinear case the nonlinearity partly compensates the Bloch-Siegert shift, which also influences the relative peak heights, as we argued in section \ref{Hamiltonian}. 

\subsection{Higher temperatures}\label{HT}
\begin{figure}[h!]
\begin{flushright}
  \resizebox{\linewidth}{!}{
  \includegraphics{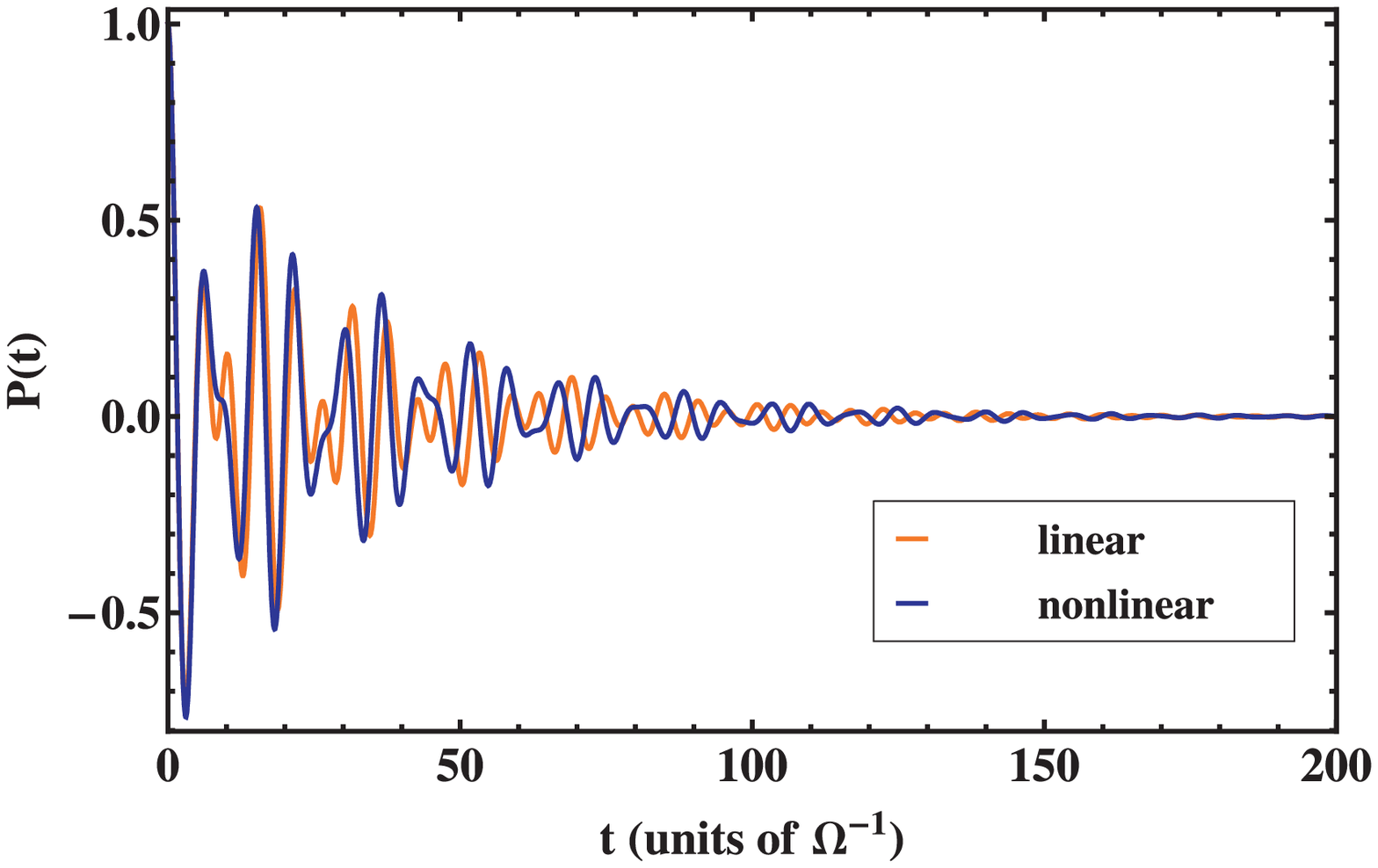}}
\end{flushright}
\begin{flushright}
  \resizebox{\linewidth}{!}{
  \includegraphics{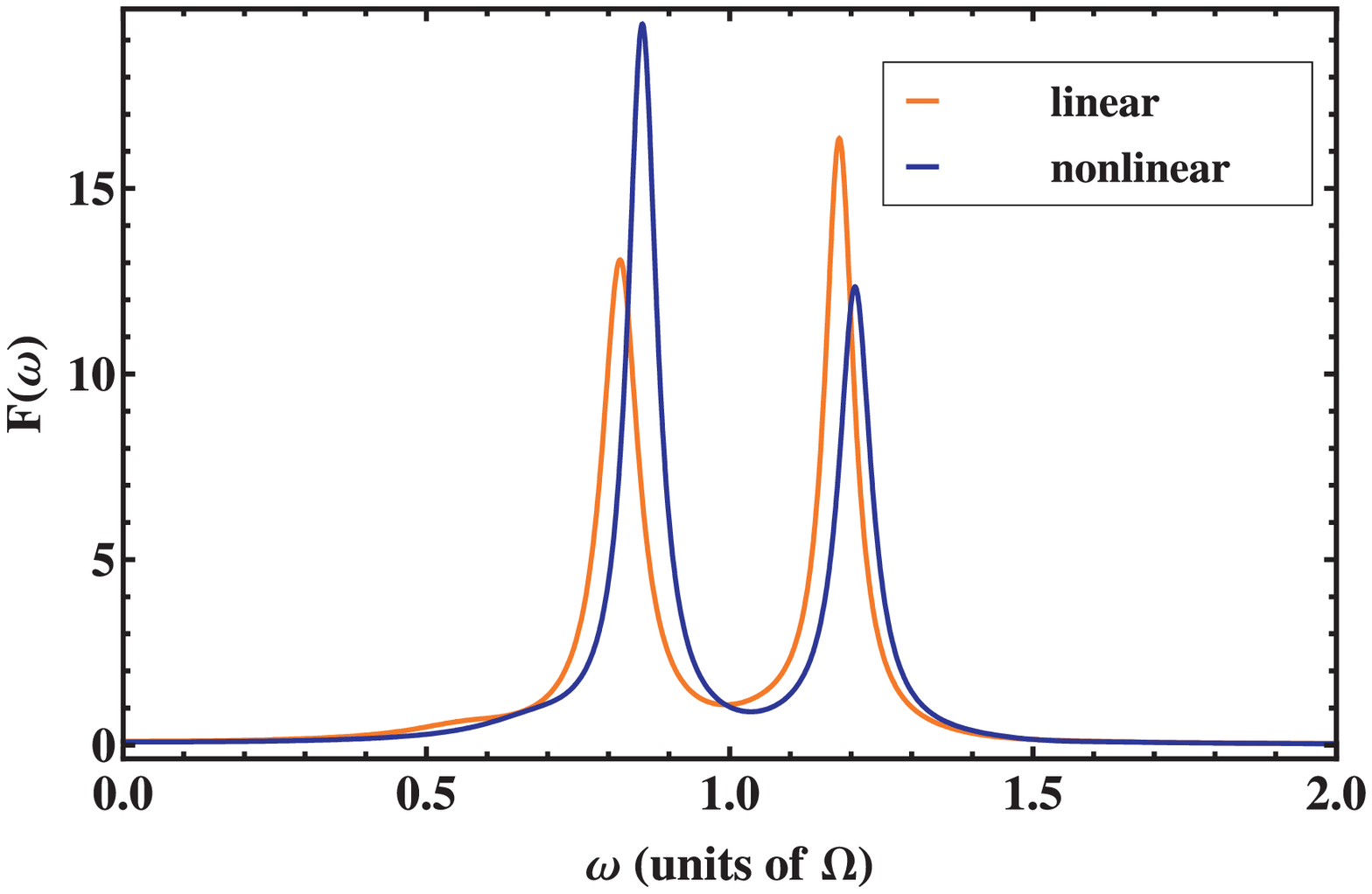}}
\end{flushright}
\caption{(Color online) $P(t)$ and its Fourier transform $F(\omega)$ for the parameters: $\alpha=0.02\hbar\Omega$, $g=0.18\Omega$, $\kappa=0.0154$, $\varepsilon=0$, $\Delta_b=\Omega$ as in figure \ref{FFig1}, but $\beta=3(\hbar\Omega)^{-1}$. For comparison we plotted the linear case in orange (light gray).\label{FFig2}  }
\end{figure}

\begin{figure}[h!]
\begin{flushright}
  \resizebox{\linewidth}{!}{
  \includegraphics{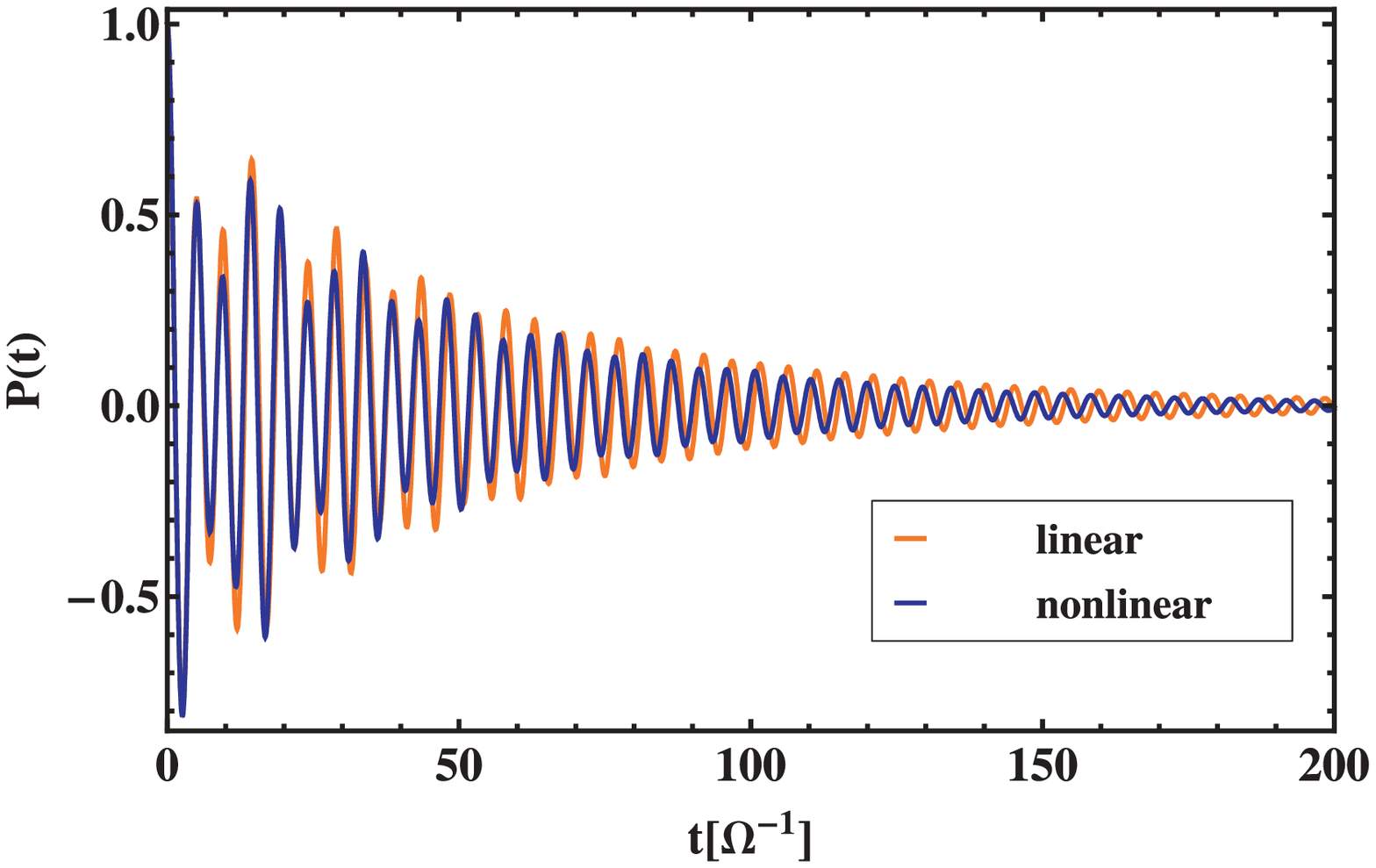}}
\end{flushright}
\begin{flushright}
  \resizebox{\linewidth}{!}{
  \includegraphics{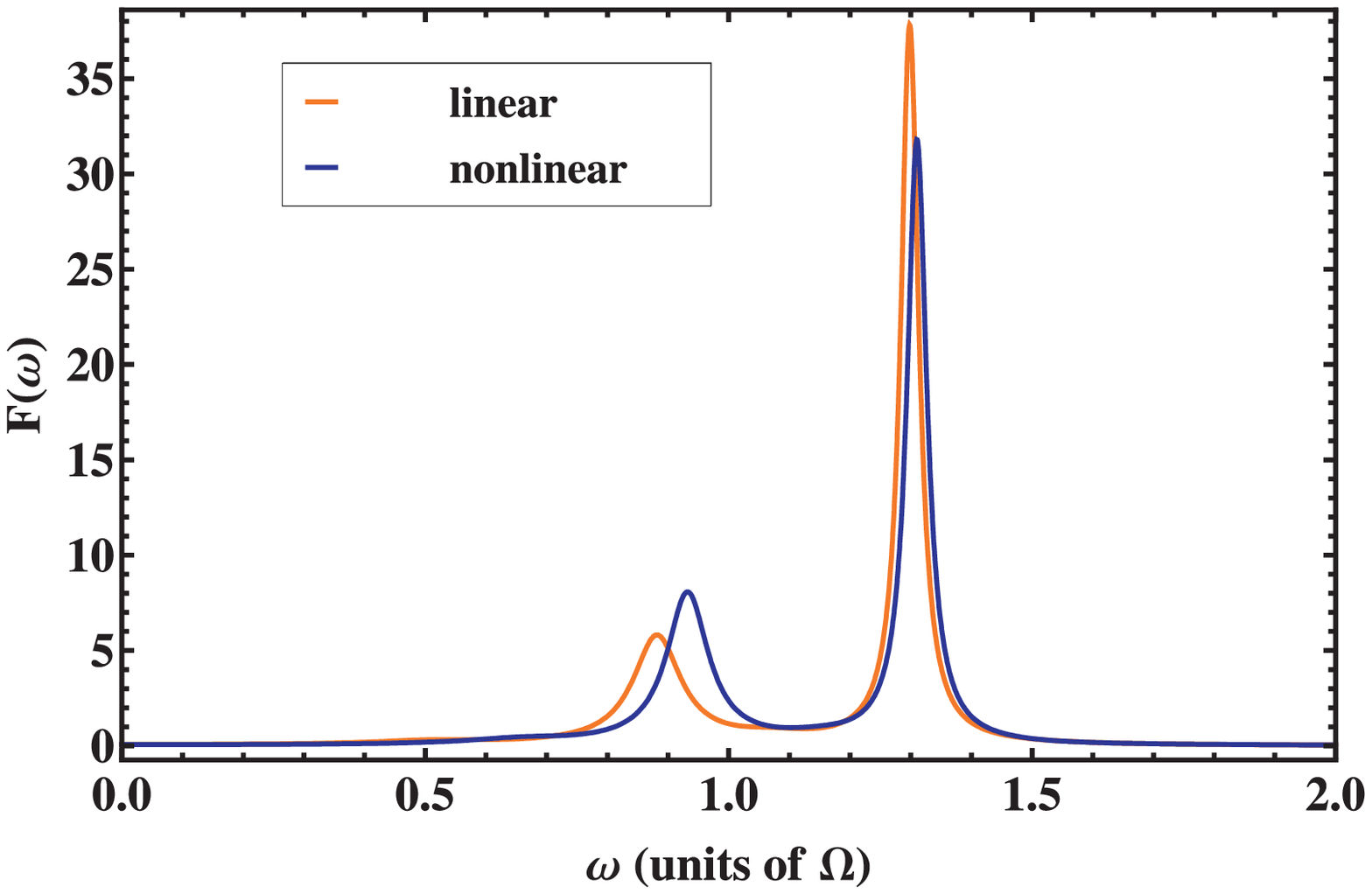}}
\end{flushright}
\caption{(Color online) $P(t)$ and its Fourier transform $F(\omega)$ for the parameters: $\alpha=0.02\hbar\Omega$, $g=0.18\Omega$, $\kappa=0.0154$, $\varepsilon=0$, $\Delta_b=1.18\Omega$, corresponding to the oscillator transition from $\ket{3}\rightarrow\ket{2}$, and $\beta=3(\hbar\Omega)^{-1}$. For comparison we plotted the linear case in orange (light gray).\label{FFig3}  }
\end{figure}

To investigate the influence of temperature we show in figure \ref{FFig2} $P( t )$ and the corresponding $F(\omega)$ for the same parameters as in figure \ref{FFig1}, but at inverse temperature: $\beta=3/(\hbar\Omega)$. 
By increasing the temperature higher oscillator levels are populated and influence the dynamics of the qubit. We calculated the corresponding equations for the long time dynamics within the SEA. The relaxation matrix $\mathcal{L}_{{\rm relax}}$ for the rate $\Gamma_{{\rm r}}$ was calculated inplementing higher levels, until $\ket{8}_{{\rm eff}}$.\\
We plot for comparison also the linear oscillator case. We observe again the overall shift of the resonance frequencies to higher values and that a new shoulder arises in the Fourier spectrum. It corresponds to the transition frequency $\omega_{32}$ (see also figure \ref{ft6} bottom). We checked numerically that the structure of the Fourier spectrum can be fully respresented by summation of the six contributions in $P(t)$ with the frequencies $\omega_{10}$, $\omega_{20}$, $\omega_{32}$, $\omega_{42}$, $\omega_{13}$, and $\omega_{14}$. These six contributions arise due to the finite populations of the involved levels. Therefore the appearance of additional shoulders in the dynamics is a pure temperature effect, which is also seen in the corresponding linear case. The frequency shift induced by the nonlinearity is much larger for the higher levels. The effect of temperature is also reflected in the height of the dominating peaks, which is decreased for higher temperatures. The temperature can not influence which peak is dominant. This means by comparing figure \ref{FFig1} with figure \ref{FFig2} that in both figures in the nonlinear case the peak corresponding to $\omega_{10}$ dominates over the one corresponding $\omega_{20}$.\\
The use of a nonlinear oscillator instead of a linear one has advantages which rely in the fact that the energy spectrum of the nonlinear oscillator is not equidistant. Supposing that the TLS frequency $\Delta_b$ can be tuned, it is in case of the nonlinear oscillator possible to have the TLS in resonance with exactly one and only one nonlinear oscillator state transition. All other transitions are then off resonance/detuned. For the linear oscillator in resonance with the TLS the number of possible transitions is in principle infinite. Therefore we determine in the following the dynamics of the qubit by putting the qubit in resonance with the nonlinear oscillator transition $\ket{3}\rightarrow\ket{2}$, see figure \ref{FFig3}. 
We read off from figure \ref{FFig3} that the detuning compared to figure \ref{FFig2} results in the enhancement of the $\omega_{20}$-peak, whereas the other dominating peak is shrinked. This is due to the different resonance conditions leading to opposite weights of the peaks for the nonlinear case in figure \ref{FFig3} compared to figure \ref{FFig2}. However a peak corresponding to higher transitions is not seen. The reason for this is the small population of the higher oscillator levels involved.%
\section{Conclusions}\label{Conclusions}
To conclude, we determined the dynamics of a TLS which is coupled via a nonlinear oscillator to an environment described by an Ohmic spectral density. We restricted ourselves to the regime of weak nonlinearity, weak damping and moderate coupling of oscillator and TLS. To diagonalize the qubit-nonlinear-oscillator Hamiltonian we used Van Vleck perturbation theory, hence avoiding the use of the rotating wave approximation (RWA). Within the RWA and for vanishing nonlinearity our model would reduce to the Jaynes-Cummings Hamiltonian. In section \ref{Hamiltonian}, an analytical expression for the non-dissipative dynamics is given, which accounts for the infinite Hilbert space of the composed system. The influence of the nonlinearity onto the qubit dynamics is determined and compared to the linear case.

At low temperatures $k_BT<\hbar\Omega,\hbar\Delta_b$ this infinite Hilbert space can be truncated such that the transition processes between the ground state and the two first excited energy levels of the qubit-nonlinear-oscillator system dominate the dynamics.
As in the linear case this yields a pronounced vacuum Rabi splitting.\\ 
To investigate the influence of the bath we solved the Bloch-Redfield Markovian master equation for the density-matrix of the qubit-nonlinear-oscillator system numerically. For an analytical treatment we considered three kinds of approximations: first a full secular approximation including a low temperature approximation, where all fast oscillating terms are neglected. Second an ansatz for the long-time dynamics allows a general expression for the relaxation and dephasing rates of the qubit. The third approximation was a partial secular approximation reproducing almost perfectly the exact numerical solution. A comparison of these three analytical approaches showed good agreement with the numerical solution. Finally, we investigated the effect of the non-equidistant energy spectrum of the nonlinear oscillator on the TLS dynamics. To do so, we allowed higher temperatures to populate higher levels and moreover we concentrated on the actual transition of the nonlinear oscillator from $\ket{3}\rightarrow\ket{2}$. We observed the rise of additional shoulders in the Fourier spectrum and showed that the shift in the transition frequencies is much larger if higher oscillator levels are involved. %
\section{acknowledgments}
We acknowledge support by the DFG under the funding programs GRK 638, SFB 631.
\newpage
\appendix
\begin{widetext}
\section{Van Vleck perturbation theory}\label{Vleck}
In our case the perturbation $\mathcal{H}_{{\rm Int}}$ is proportional to $B+B^\dagger$. Therefore we consider first the action of this operator on arbitrary nonlinear oscillator states $\ket{l},\ket{m}$:
\begin{eqnarray}
\bra{l}B+B^\dagger\ket{m}&=&\bra{l}\left[\sqrt{m}\ket{m-1}_0+a_2^{(m)}\sqrt{m+2}\ket{m+1}_0+
 a_{-2}^{(m)}\sqrt{m-2}\ket{m-3}_0+ a_{-4}^{(m)}\sqrt{m-4}\ket{m-5}_0\right.\\
&&\nonumber+a_{4}^{(m)}\sqrt{m+4}\ket{m+3}_0+\sqrt{m+1}\ket{m+1}_0+a_2^{(m)}\sqrt{m+3}\ket{m+3}_0+a_{-2}^{(m)}\sqrt{m-1}\ket{m-1}_0\\
&&\left.\nonumber+a_{-4}^{(m)}\sqrt{m-3}\ket{m-3}_0+a_{4}^{(m)}\sqrt{m+5}\ket{m+5}_0
\right]+\mathcal{O}(\alpha^2),
\end{eqnarray}
where $\ket{l}_0$ denotes an eigenstates of the linear oscillator.
Now we have different cases:
\begin{eqnarray}
l=m-1&:&\quad\bra{m-1}(B+B^\dagger)\ket{m}=\sqrt{m}+a_{-2}^{(m)}\sqrt{m-1}+a_2^{(m-1)}\sqrt{m+1}+\mathcal{O}(\alpha^2),\\
l=m-3&:&\quad\bra{m-3}(B+B^\dagger)\ket{m}=a_{-2}^{(m)}\sqrt{m-2}+a_{-4}^{(m)}\sqrt{m-3}+a_2^{(m-3)}\sqrt{m}+a_4^{(m-3)}\sqrt{m+1}+\mathcal{O}(\alpha^2),\nonumber\\
l=m-5&:&\quad\bra{m-5}(B+B^\dagger)\ket{m}=a_{-4}^{(m)}\sqrt{m-4}+a_4^{(m-5)}\sqrt{m}+\mathcal{O}(\alpha^2)=0,\nonumber\\
l=m+1&:&\quad\bra{m+1}(B+B^\dagger)\ket{m}=\sqrt{m+1}+a_{-2}^{(m+1)}\sqrt{m}+a_2^{(m)}\sqrt{m+2}+\mathcal{O}(\alpha^2),\nonumber\\
l=m+3&:&\quad
\bra{m+3}(B+B^\dagger)\ket{m}=a_{-2}^{(m+3)}\sqrt{m+1}+a_{-4}^{(m+3)}\sqrt{m}+a_2^{(m)}\sqrt{m+3}+a_4^{(m)}\sqrt{m+4}+\mathcal{O}(\alpha^2),\nonumber\\
l=m+5&:&
\bra{m+5}(B+B^\dagger)\ket{m}=a_{-4}^{(m+5)}\sqrt{m+1}+a_4^{(m)}\sqrt{m+5}+\mathcal{O}(\alpha^2)=0.\nonumber
\end{eqnarray}
Due to the manifold structure we only have to consider for Van Vleck perturbation theory the matrix elements involving $l=m\pm1$, $l=m\pm3$. Therefore we introduce the notations:
\begin{eqnarray}
n_1(j)&=&
\sqrt{j+1} \left(1+\frac{\sqrt{j}a_{-2}^{(j+1)}}{\sqrt{j+1}}+\frac{a_2^{(j)}
   \sqrt{j+2}}{\sqrt{j+1}}\right) = \sqrt{j+1}\left[1-\frac{3}{2\hbar\Omega}\alpha(j+1)\right] ,\\
n_3(j,\alpha)&=&a_{-2}^{(j)}\sqrt{j-2}+a_{-4}^{(j)}\sqrt{j-3}+a_2^{(j-3)}\sqrt{j}+a_4^{(j-3)}\sqrt{j+1}=\frac{\alpha}{4\hbar\Omega}\sqrt{j(j-1)(j-2)}.
\end{eqnarray}
The non-vanishing matrix elements for the transformation matrix are in first order:
\begin{equation}
iS^{(1)}_{(j-1){\rm e},j {\rm e}}=\frac{\langle {\rm e},j-1|\mathcal{H}_{\rm Int}|{\rm e},j\rangle}{E_{{\rm e}(j-1)}-E_{{\rm e}j}}%
=\frac{g\frac{\varepsilon}{\Delta_b}n_1(j-1)}{\Omega+\frac{3}{2\hbar}\alpha\cdot 2j}
=\frac{g \varepsilon  \sqrt{j}}{\Delta_b \Omega}\left[1-\frac{9}{2\hbar\Omega}\alpha j\right]+\mathcal{O}(\alpha^2),
\end{equation}
\begin{equation}
iS^{(1)}_{j {\rm g},(j+1){\rm g}}=\frac{\langle {\rm g},j|\mathcal{H}_{\rm Int}|{\rm g},j+1\rangle}{E_{{\rm g}j}-E_{{\rm g}(j+1)}}
=-\frac{g\frac{\varepsilon}{\Delta_b}n_1(j)}{\Omega+\frac{3}{2\hbar}\alpha\cdot 2(j+1)}
=-\frac{g \varepsilon  \sqrt{j+1}}{\Delta_b \Omega}\left[1-\frac{9}{2\hbar\Omega}\alpha (j+1)\right]+\mathcal{O}(\alpha^2),\nonumber
\end{equation}
\begin{equation}
iS^{(1)}_{j{\rm g},(j+1){\rm e}}=\frac{\langle {\rm g},j|\mathcal{H}_{\rm Int}|{\rm e},j+1\rangle}{E_{{\rm g}j}-E_{{\rm e}(j+1)}}%
=\frac{g\frac{\Delta_0}{\Delta_b}n_1(j)}{\Delta_b+\Omega+\frac{3}{2\hbar}\alpha\cdot 2(j+1)}
=\frac{g\Delta_0\sqrt{j+1}}{\Delta_b(\Delta_b+\Omega)}\left[1-\frac{3\alpha(j+1)(\Delta_b+3\Omega)}{2\hbar\Omega(\Delta_b+\Omega)}\right]+\mathcal{O}(\alpha^2),\nonumber
\end{equation}
\begin{equation}
iS^{(1)}_{j{\rm e},(j+3) {\rm e}}=\frac{\langle {\rm e},j|\mathcal{H}_{\rm Int}|{\rm e},j+3\rangle}{E_{{\rm e}j}-E_{{\rm e}(j+3)}}%
=\frac{g\frac{\varepsilon}{\Delta_b}n_3(j+3,\alpha)}{3\Omega}+\mathcal{O}(\alpha^2),\nonumber
\end{equation}
\begin{equation}
iS^{(1)}_{j {\rm g},(j+3){\rm g}}=\frac{\langle {\rm g},j|\mathcal{H}_{\rm Int}|{\rm g},j+3\rangle}{E_{{\rm g}j}-E_{{\rm g}(j+3)}}
=-\frac{g\frac{\varepsilon}{\Delta_b}n_3(j+3,\alpha)}{3\Omega}+\mathcal{O}(\alpha^2),\nonumber
\end{equation}
\begin{equation}
iS^{(1)}_{j{\rm g},(j+3){\rm e}}=\frac{\langle {\rm g},j|\mathcal{H}_{\rm Int}|{\rm e},j+3\rangle}{E_{{\rm g}j}-E_{{\rm e}(j+3)}}
=\frac{g\frac{\Delta_0}{\Delta_b}n_3(j+3,\alpha)}{\Delta_b+3\Omega}+\mathcal{O}(\alpha^2),\nonumber
\end{equation}
\begin{equation}
iS^{(1)}_{j{\rm e},(j+3){\rm g}}=\frac{\langle {\rm e},j|\mathcal{H}_{\rm Int}|{\rm g},j+3\rangle}{E_{{\rm e}j}-E_{{\rm g}(j+3)}}=\frac{g\frac{\Delta_0}{\Delta_b}n_3(j+3,\alpha)}{-\Delta_b+3\Omega}+\mathcal{O}(\alpha^2).\nonumber
\end{equation}
Due to the fact that $n_3(j,\alpha)$ is a purely nonlinear contribution, we can reduce the possible contributions for the second order of the transformation matrix. Restricting to lowest order in the nonlinearity the non-vanishing contributions are either combinations of involving twice $n_1(j)$ and expanding this afterwards to first order in the nonlinearity or combinations of both $n_1(j)$ and $n_3(j,\alpha)$, while $n_1(j)$ is reduced in this case to the zeroth order in the nonlinearity, because $n_3(j,\alpha)$ is already of first order in the nonlinearity.
For the second order we obtain:
\begin{eqnarray}
iS^{(2)}_{j{\rm e},(j+2){\rm g}}&=&\frac{1}{E_{{\rm g}(j+2)}-E_{{\rm e}j}}\left[
\frac{\bra{{\rm e},j}\mathcal{H}_{\rm Int}\ket{{\rm e},j+1}\bra{{\rm e},j+1}\mathcal{H}_{\rm Int}\ket{{\rm g},j+2} }{E_{{\rm e}(j+1)}-E_{{\rm e}j}}+
\frac{\bra{{\rm e},j}\mathcal{H}_{\rm Int}\ket{{\rm g},j+1}\bra{{\rm g},j+1}\mathcal{H}_{\rm Int}\ket{{\rm g},j+2}}{E_{{\rm g}(j+1)}-E_{{\rm g}(j+2)}}
\right]\nonumber\\
&&+\frac{\bra{{\rm e},j}\mathcal{H}_{\rm Int}\ket{{\rm e},j-1}\bra{{\rm e},j-1}\mathcal{H}_{\rm Int}\ket{{\rm g},j+2}}{2(E_{{\rm g}(j+2)}-E_{{\rm e}j})}\left[\frac{1}{E_{{\rm e}(j-1)}-E_{{\rm e}j}}+\frac{1}{E_{{\rm e}(j-1)}-E_{{\rm g}(j+2)}}\right]\nonumber\\
&&+\frac{\bra{{\rm e},j}\mathcal{H}_{\rm Int}\ket{{\rm g},j-1}\bra{{\rm g},j-1}\mathcal{H}_{\rm Int}\ket{{\rm g},j+2}}{2(E_{{\rm g}(j+2)}-E_{{\rm e}j})}\left[\frac{1}{E_{{\rm g}(j-1)}-E_{{\rm e}j}}+\frac{1}{E_{{\rm g}(j-1)}-E_{{\rm g}(j+2)}}\right]\nonumber\\
&&+\frac{\bra{{\rm e},j}\mathcal{H}_{\rm Int}\ket{{\rm g},j+3}\bra{{\rm g},j+3}\mathcal{H}_{\rm Int}\ket{{\rm g},j+2}}{2(E_{{\rm g}(j+2)}-E_{{\rm e}j})}\left[\frac{1}{E_{{\rm g}(j+3)}-E_{{\rm e}j}}+\frac{1}{E_{{\rm g}(j+3)}-E_{{\rm g}(j+2)}}\right]\nonumber\\
&&+\frac{\bra{{\rm e},j}\mathcal{H}_{\rm Int}\ket{{\rm e},j+3}\bra{{\rm e},j+3}\mathcal{H}_{\rm Int}\ket{{\rm g},j+2}}{2(E_{{\rm g}(j+2)}-E_{{\rm e}j})}\left[\frac{1}{E_{{\rm e}(j+3)}-E_{{\rm e}j}}+\frac{1}{E_{{\rm e}(j+3)}-E_{{\rm g}(j+2)}}\right]\nonumber\\
&=&\frac{2\hbar^2 g^2\varepsilon\Delta_0\sqrt{(j+1)(j+2)}}{\Delta_b^2\hbar^2\Omega(2\Omega-\Delta_b)}\left[1+\frac{3\alpha(2j+3)(\Delta_b-3\Omega)}{\hbar\Omega(2\Omega-\Delta_b)}\right]\nonumber\\&&
+\frac{g^2 (2 j+3) \sqrt{(j+1)(j+2)} \alpha  \Delta_0 \varepsilon  (\Delta_b -5 \Omega)}{12 \hbar \Omega^2 \left(\Delta_b ^4-4
   \Omega \Delta_b ^3+\Omega^2 \Delta_b ^2+6 \Omega^3 \Delta_b \right)}+\mathcal{O}(\alpha^2),\nonumber
\end{eqnarray}
\begin{eqnarray}
iS^{(2)}_{j{\rm g},(j+2){\rm g}}&=&\frac{\bra{{\rm g},j}\mathcal{H}_{\rm Int}\ket{{\rm g},j+1}\bra{{\rm g},j+1}\mathcal{H}_{\rm Int}\ket{{\rm g},j+2}}{2(E_{{\rm g}(j+2)}-E_{{\rm g}j})}\left[\frac{1}{E_{{\rm g}(j+1)}-E_{{\rm g}j}}+\frac{1}{E_{{\rm g}(j+1)}-E_{{\rm g}(j+2)}}\right]\nonumber\\
&&+\frac{\bra{{\rm g},j}\mathcal{H}_{\rm Int}\ket{{\rm e},j+1}\bra{{\rm e},j+1}\mathcal{H}_{\rm Int}\ket{{\rm g},j+2} }{(E_{{\rm g}(j+2)}-E_{{\rm g}j})(E_{{\rm e}(j+1)}-E_{{\rm g}j})}\nonumber\\
&&+\frac{\bra{{\rm g},j}\mathcal{H}_{\rm Int}\ket{{\rm g},j-1}\bra{{\rm g},j-1}\mathcal{H}_{\rm Int}\ket{{\rm g},j+2}}{2(E_{{\rm g}(j+2)}-E_{{\rm g}j})}\left[\frac{1}{E_{{\rm g}(j-1)}-E_{{\rm g}j}}+\frac{1}{E_{{\rm g}(j-1)}-E_{{\rm g}(j+2)}}\right]\nonumber\\
&&+\frac{\bra{{\rm g},j}\mathcal{H}_{\rm Int}\ket{{\rm g},j+3}\bra{{\rm g},j+3}\mathcal{H}_{\rm Int}\ket{{\rm g},j+2}}{2(E_{{\rm g}(j+2)}-E_{{\rm g}j})}\left[\frac{1}{E_{{\rm g}(j+3)}-E_{{\rm g}j}}+\frac{1}{E_{{\rm g}(j+3)}-E_{{\rm g}(j+2)}}\right]\nonumber\\
&&+\frac{\bra{{\rm g},j}\mathcal{H}_{\rm Int}\ket{{\rm e},j+3}\bra{{\rm e},j+3}\mathcal{H}_{\rm Int}\ket{{\rm g},j+2}}{2(E_{{\rm g}(j+2)}-E_{{\rm g}j})}\left[\frac{1}{E_{{\rm e}(j+3)}-E_{{\rm g}j}}+\frac{1}{E_{{\rm e}(j+3)}-E_{{\rm g}(j+2)}}\right]\nonumber\\
&&+\frac{\bra{{\rm g},j}\mathcal{H}_{\rm Int}\ket{{\rm e},(j-1)}\bra{{\rm e},j-1}\mathcal{H}_{\rm Int}\ket{{\rm g},(j+2)} }{(E_{{\rm g}(j+2)}-E_{{\rm g}j})(E_{{\rm e}(j-1)}-E_{{\rm g}(j+2)})}\nonumber\\
&=&\frac{\hbar g^2  \sqrt{(j+1)(j+2)} }{2 \Delta_b ^2 \Omega}\left[\frac{3\alpha\varepsilon^2}{2\hbar^2\Omega^2}+\frac{\Delta_0^2}{\hbar(\Delta_b+\Omega)}\left(1-\frac{3\alpha((2j+3)\Delta_b+\Omega(3j+4))}{\hbar(\Delta_b+\Omega)\Omega}\right)\right]\nonumber\\&&
+\frac{ g^2 \sqrt{(j+1)(j+2)} \alpha}{8 \hbar \Delta_b^2 \Omega^2}  \left(\frac{2\varepsilon^2}{\Omega}+\frac{ \left((2 j+3) \Delta_b ^2+3 (j-1) \Omega \Delta_b -3 (j+6) \Omega^2\right) \Delta_0^2}{(\Delta_b -3 \Omega) \left(\Delta_b ^2+4 \Omega \Delta_b +3 \Omega^2\right)}\right)+\mathcal{O}(\alpha^2),\nonumber
\end{eqnarray}

\begin{eqnarray}
iS^{(2)}_{j{\rm e},(j+2){\rm e}}&=&\frac{\bra{{\rm e},j}\mathcal{H}_{\rm Int}\ket{{\rm e},j+1}\bra{{\rm e},j+1}\mathcal{H}_{\rm Int}\ket{{\rm e},j+2}}{2(E_{{\rm e}(j+2)}-E_{{\rm e}j})}\left[\frac{1}{E_{{\rm e}(j+1)}-E_{{\rm e}j}}+\frac{1}{E_{{\rm e}(j+1)}-E_{{\rm e}(j+2)}}\right]\nonumber\\
&&+\frac{\bra{{\rm e},j}\mathcal{H}_{\rm Int}\ket{{\rm g},j+1}\bra{{\rm g},j+1}\mathcal{H}_{\rm Int}\ket{{\rm e},j+2} }{(E_{{\rm e}(j+2)}-E_{{\rm e}j})(E_{{\rm g}(j+1)}-E_{{\rm e}(j+2)})}\nonumber\\
&&+\frac{\bra{{\rm e},j}\mathcal{H}_{\rm Int}\ket{{\rm g},j-1}\bra{{\rm g},j-1}\mathcal{H}_{\rm Int}\ket{{\rm e},j+2}}{2(E_{{\rm e}(j+2)}-E_{{\rm e}j})}\left[\frac{1}{E_{{\rm g}(j-1)}-E_{{\rm e}j}}+\frac{1}{E_{{\rm g}(j-1)}-E_{{\rm e}(j+2)}}\right]\nonumber\\
&&+\frac{\bra{{\rm e},j}\mathcal{H}_{\rm Int}\ket{{\rm e},j-1}\bra{{\rm e},j-1}\mathcal{H}_{\rm Int}\ket{{\rm e},j+2}}{2(E_{{\rm e}(j+2)}-E_{{\rm e}j})}\left[\frac{1}{E_{{\rm e}(j-1)}-E_{{\rm e}j}}+\frac{1}{E_{{\rm e}(j-1)}-E_{{\rm g}(j+2)}}\right]\nonumber\\
&&+\frac{\bra{{\rm e},j}\mathcal{H}_{\rm Int}\ket{{\rm e},j+3}\bra{{\rm e},j+3}\mathcal{H}_{\rm Int}\ket{{\rm e},j+2}}{2(E_{{\rm e}(j+2)}-E_{{\rm e}j})}\left[\frac{1}{E_{{\rm e}(j+3)}-E_{{\rm g}j}}+\frac{1}{E_{{\rm e}(j+3)}-E_{{\rm e}(j+2)}}\right]\nonumber\\
&&+\frac{\bra{{\rm e}j}\mathcal{H}_{\rm Int}\ket{{\rm g},j+3}\bra{{\rm g},j+3}\mathcal{H}_{\rm Int}\ket{{\rm e},j+2} }{(E_{{\rm e}(j+2)}-E_{{\rm e}j})(E_{{\rm g}(j+3)}-E_{{\rm e}j})}\nonumber\\
&=&\frac{\hbar^2 g^2\sqrt{(j+1)(j+2)}}{\Delta_b^2 2\hbar\Omega}\left[-\frac{\Delta_0^2}{\hbar(\Omega+\Delta_b)}\left(1-\frac{3\alpha(\Delta_b(2j+3)+\Omega(3j+5))}{\hbar(\Delta_b+\Omega)\Omega}\right)+\frac{3\alpha\varepsilon^2}{\hbar^2\Omega^2}\right]
\nonumber\\&&
+\frac{g^2 \alpha\sqrt{(j+1)(j+2)}}{8 \hbar \Delta_b ^2 \Omega^2}  \left(-\frac{(3+2j)\Delta_b^2+3(4+j)\Delta_b\Omega-3(j-3)\Omega^2}{\Delta_b ^3+\Delta_b^2\Omega-9\Omega^2 \Delta_b -9 \Omega^3}\right.\left.\right.\nonumber\\&&\left.
+\frac{2\varepsilon^2}{\Omega}\right)+\mathcal{O}\left(\alpha ^2\right),\nonumber
\end{eqnarray}
\begin{eqnarray}
iS^{(2)}_{j{\rm g},j{\rm e}}&=&\frac{\bra{{\rm g},j}\mathcal{H}_{\rm Int}\ket{{\rm g},j-1}\bra{{\rm g},j-1}\mathcal{H}_{\rm Int}\ket{{\rm e},j}}{2(E_{{\rm e}j}-E_{{\rm g}j})}\left[\frac{1}{E_{{\rm g}(j-1)}-E_{{\rm g}j}}+\frac{1}{E_{{\rm g}(j-1)}-E_{{\rm e}j}}\right]\nonumber\\
&&+\frac{\bra{{\rm g},j}\mathcal{H}_{\rm Int}\ket{{\rm e},j+1}\bra{{\rm e},j+1}\mathcal{H}_{\rm Int}\ket{{\rm e},j}}{2(E_{{\rm e}j}-E_{{\rm g}j})}\left[\frac{1}{E_{{\rm e}(j+1)}-E_{{\rm g}j}}+\frac{1}{E_{{\rm e}(j+1)}-E_{{\rm e}j}}\right]\nonumber\\
&&+\frac{\bra{{\rm g},j}\mathcal{H}_{\rm Int}\ket{{\rm e},j-1}\bra{{\rm e},j-1}\mathcal{H}_{\rm Int}\ket{{\rm e},j} }{(E_{{\rm e}j}-E_{{\rm g}j})(E_{{\rm e}(j-1)}-E_{{\rm e}j})}
+\frac{\bra{{\rm g},j}\mathcal{H}_{\rm Int}\ket{{\rm g},j+1}\bra{{\rm g},j+1}\mathcal{H}_{\rm Int}\ket{{\rm e},j} }{(E_{{\rm e}j}-E_{{\rm g}j})(E_{{\rm g}(j+1)}-E_{{\rm g}j})}\nonumber\\
&=&\frac{g^2\Delta_0\varepsilon}{\hbar\Delta_b^2\Omega(\Delta_b+\Omega)}\left[-\frac{\hbar(2j+1)}{2}+\frac{3\alpha(2j^2+2j+1)(2\Delta_b+3\Omega)}{2\Omega(\Delta_b+\Omega)}\right]
+\mathcal{O}\left(\alpha ^2\right)\nonumber,
\end{eqnarray}

\begin{eqnarray}
iS^{(2)}_{j{\rm g},(j+2) {\rm e}}&=&\frac{\bra{{\rm g},j}\mathcal{H}_{\rm Int}\ket{{\rm g},j+1}\bra{{\rm g},j+1}\mathcal{H}_{\rm Int}\ket{{\rm e},j+2}}{2(E_{{\rm e}(j+2)}-E_{{\rm g}j})}\left[\frac{1}{E_{{\rm g}(j+1)}-E_{{\rm g}j}}+\frac{1}{E_{{\rm g}(j+1)}-E_{{\rm e}(j+2)}}\right]\nonumber\\
&&+\frac{\bra{{\rm g},j}\mathcal{H}_{\rm Int}\ket{{\rm e},j+1}\bra{{\rm e},j+1}\mathcal{H}_{\rm Int}\ket{{\rm e},j+2}}{2(E_{{\rm e}(j+2)}-E_{{\rm g}j})}\left[\frac{1}{E_{{\rm e}(j+1)}-E_{{\rm g}j}}+\frac{1}{E_{{\rm e}(j+1)}-E_{{\rm e}(j+2)}}\right]\nonumber\\
&&+\frac{\bra{{\rm g},j}\mathcal{H}_{\rm Int}\ket{{\rm g},j-1}\bra{{\rm g},j-1}\mathcal{H}_{\rm Int}\ket{{\rm e},j+2}}{2(E_{{\rm e}(j+2)}-E_{{\rm g}j})}\left[\frac{1}{E_{{\rm g}(j-1)}-E_{{\rm g}j}}+\frac{1}{E_{{\rm g}(j-1)}-E_{{\rm e}(j+2)}}\right]\nonumber\\
&&+\frac{\bra{{\rm g},j}\mathcal{H}_{\rm Int}\ket{{\rm e},j+3}\bra{{\rm e},j+3}\mathcal{H}_{\rm Int}\ket{{\rm e},j+2}}{2(E_{{\rm e}(j+2)}-E_{{\rm g}j})}\left[\frac{1}{E_{{\rm e}(j+3)}-E_{{\rm g}j}}+\frac{1}{E_{{\rm e}(j+3)}-E_{{\rm e}(j+2)}}\right]\nonumber\\
&&+\frac{\bra{{\rm g},j}\mathcal{H}_{\rm Int}\ket{{\rm e},j-1}\bra{{\rm e},j-1}\mathcal{H}_{\rm Int}\ket{{\rm e},j+2}}{(E_{{\rm e}(j+2)}-E_{{\rm g}j})(E_{{\rm e}(j-1)}-E_{{\rm e}(j+2)})}
+\frac{\bra{{\rm g},j}\mathcal{H}_{\rm Int}\ket{{\rm g},j+3}\bra{{\rm g},j+3}\mathcal{H}_{\rm Int}\ket{{\rm e},j+2}}{(E_{{\rm e}(j+2)}-E_{{\rm g}j})(E_{{\rm g}(j+3)}-E_{{\rm g}j})}\nonumber\\
&=&\frac{\hbar^2 g^2\varepsilon\Delta_0\sqrt{(j+1)(j+2)}}{2\Delta_b^2\hbar(2\Omega+\Delta_b)}\left[-\frac{2\Delta_b}{\hbar\Omega(\Omega+\Delta_b)}+\frac{3\alpha\Delta_b(2j+3)(2\Delta_b^2+9\Delta_b\Omega+8\Omega^2)}{\hbar^2\Omega^2(\Omega+\Delta_b)^2(2\Omega+\Delta_b)}\right]\nonumber\\&&
+\frac{g^2 (2 j+3) \sqrt{(j+1)(j+2)} \alpha  \Delta_0 \varepsilon  (\Delta_b +6 \Omega)}{24 \hbar \Delta_b ^2 \Omega^2 \left(\Delta_b
   ^2+5 \Omega \Delta_b +6 \Omega^2\right)}+\mathcal{O}\left(\alpha ^2\right)\nonumber,
\end{eqnarray}

\begin{eqnarray}
iS^{(2)}_{j{\rm g},(j+4) {\rm g}}&=&\frac{\bra{{\rm g},j}\mathcal{H}_{\rm Int}\ket{{\rm g},j+1}\bra{{\rm g},j+1}\mathcal{H}_{\rm Int}\ket{{\rm g},j+4}}{2(E_{{\rm g}(j+4)}-E_{{\rm g}j})}\left[\frac{1}{E_{{\rm g}(j+1)}-E_{{\rm g}j}}+\frac{1}{E_{{\rm g}(j+1)}-E_{{\rm g}(j+4)}}\right]\nonumber\\
&&+\frac{\bra{{\rm g},j}\mathcal{H}_{\rm Int}\ket{{\rm e},j+1}\bra{{\rm e},j+1}\mathcal{H}_{\rm Int}\ket{{\rm g},(j+4)}}{2(E_{{\rm g}(j+4)}-E_{{\rm g}j})}\left[\frac{1}{E_{{\rm e}(j+1)}-E_{{\rm g}j}}+\frac{1}{E_{{\rm e}(j+1)}-E_{{\rm g}(j+4)}}\right]\nonumber\\
&&+\frac{\bra{{\rm g},j}\mathcal{H}_{\rm Int}\ket{{\rm g},j+3}\bra{{\rm g},j+3}\mathcal{H}_{\rm Int}\ket{{\rm g},(j+4)}}{2(E_{{\rm g}(j+4)}-E_{{\rm g}j})}\left[\frac{1}{E_{{\rm g}(j+3)}-E_{{\rm g}j}}+\frac{1}{E_{{\rm g}(j+3)}-E_{{\rm g}(j+4)}}\right]\nonumber\\
&&+\frac{\bra{{\rm g},j}\mathcal{H}_{\rm Int}\ket{{\rm e},j+3}\bra{{\rm e},j+3}\mathcal{H}_{\rm Int}\ket{{\rm g},j+4}}{(E_{{\rm g}(j+4)}-E_{{\rm g}j})(E_{{\rm e}(j+3)}-E_{{\rm g}j})}\nonumber\\
&=&\frac{ g^2 \sqrt{(j+1)(j+2)(j+3)(j+4)} \alpha  \Delta_0^2 \left(\Delta_b ^2-3 \Omega^2\right)}{8 \hbar \Delta_b ^2 \Omega^2 \left(\Delta_b ^3+\Omega \Delta_b ^2-9 \Omega^2 \Delta_b -9 \Omega^3\right)}+\mathcal{O}(\alpha^2),\nonumber
\end{eqnarray}

\begin{eqnarray}
iS^{(2)}_{j{\rm g},(j+4) {\rm e}}&=&\frac{\bra{{\rm g},j}\mathcal{H}_{\rm Int}\ket{{\rm g},j+1}\bra{{\rm g},j+1}\mathcal{H}_{\rm Int}\ket{{\rm e},j+4}}{2(E_{{\rm e}(j+4)}-E_{{\rm g}j})}\left[\frac{1}{E_{{\rm g}(j+1)}-E_{{\rm g}j}}+\frac{1}{E_{{\rm g}(j+1)}-E_{{\rm e}(j+4)}}\right]\nonumber\\
&&+\frac{\bra{{\rm g},j}\mathcal{H}_{\rm Int}\ket{{\rm e},j+1}\bra{{\rm e},j+1}\mathcal{H}_{\rm Int}\ket{{\rm e},j+4}}{2(E_{{\rm e}(j+4)}-E_{{\rm g}j})}\left[\frac{1}{E_{{\rm e}(j+1)}-E_{{\rm g}j}}+\frac{1}{E_{{\rm e}(j+1)}-E_{{\rm e}(j+4)}}\right]\nonumber\\
&&+\frac{\bra{{\rm g},j}\mathcal{H}_{\rm Int}\ket{{\rm g},j+3}\bra{{\rm g},j+3}\mathcal{H}_{\rm Int}\ket{{\rm e},j+4}}{2(E_{{\rm e}(j+4)}-E_{{\rm g}j})}\left[\frac{1}{E_{{\rm g}(j+3)}-E_{{\rm g}j}}+\frac{1}{E_{{\rm g}(j+3)}-E_{{\rm e}(j+4)}}\right]\nonumber\\
&&+\frac{\bra{{\rm g},j}\mathcal{H}_{\rm Int}\ket{{\rm e},j+3}\bra{{\rm e},j+3}\mathcal{H}_{\rm Int}\ket{{\rm e},j+4}}{2(E_{{\rm e}(j+4)}-E_{{\rm g}j})}\left[\frac{1}{E_{{\rm e}(j+3)}-E_{{\rm g}j}}+\frac{1}{E_{{\rm e}(j+3)}-E_{{\rm e}(j+4)}}\right]\nonumber\\
&=&-\frac{g^2 \sqrt{(j+1)(j+2)(j+3)(j+4)} \alpha  \Delta_0 \varepsilon  (2 \Delta_b +5 \Omega)}{6 \hbar \Omega^2 \left(\Delta_b
   ^4+8 \Omega \Delta_b ^3+19 \Omega^2 \Delta_b ^2+12 \Omega^3 \Delta_b \right)}+\mathcal{O}(\alpha^2)\nonumber,
\end{eqnarray}

\begin{eqnarray}
iS^{(2)}_{j{\rm e},(j+4) {\rm g}}&=&\frac{\bra{{\rm e},j}\mathcal{H}_{\rm Int}\ket{{\rm e},j+1}\bra{{\rm e},j+1}\mathcal{H}_{\rm Int}\ket{{\rm g},j+4}}{2(E_{{\rm g}(j+4)}-E_{{\rm e}j})}\left[\frac{1}{E_{{\rm e}(j+1)}-E_{{\rm e}j}}+\frac{1}{E_{{\rm e}(j+1)}-E_{{\rm g}(j+4)}}\right]\nonumber\\
&&+\frac{\bra{{\rm e},j}\mathcal{H}_{\rm Int}\ket{{\rm g},j+3}\bra{{\rm g},j+3}\mathcal{H}_{\rm Int}\ket{{\rm g},j+4}}{2(E_{{\rm g}(j+4)}-E_{{\rm e}j})}\left[\frac{1}{E_{{\rm g}(j+3)}-E_{{\rm e}j}}+\frac{1}{E_{{\rm g}(j+3)}-E_{{\rm g}(j+4)}}\right]\nonumber\\
&&+\frac{\bra{{\rm e},j}\mathcal{H}_{\rm Int}\ket{{\rm e},j+3}\bra{{\rm e},j+3}\mathcal{H}_{\rm Int}\ket{{\rm g},j+4}}{(E_{{\rm g}(j+4)}-E_{{\rm e}j})(E_{{\rm e}(j+3)}-E_{{\rm e}j})}
+\frac{\bra{{\rm e},j}\mathcal{H}_{\rm Int}\ket{{\rm g},j+1}\bra{{\rm g},j+1}\mathcal{H}_{\rm Int}\ket{{\rm g},j+4}}{(E_{{\rm g}(j+4)}-E_{{\rm e}j})(E_{{\rm g}(j+1)}-E_{{\rm g}(j+4)})}\nonumber\\
&=&-\frac{g^2 \sqrt{(j+1)(j+2)(j+3)(j+4)} \alpha  \Delta_0 \varepsilon  (5 \Delta_b -12 \Omega)}{12 \hbar \Delta_b ^2 \Omega^2
   \left(\Delta_b ^2-7 \Omega \Delta_b +12 \Omega^2\right)}+\mathcal{O}(\alpha^2)\nonumber,
\end{eqnarray}

\begin{eqnarray}
iS^{(2)}_{j{\rm e},(j+4) {\rm e}}&=&\frac{\bra{{\rm e},j}\mathcal{H}_{\rm Int}\ket{{\rm e},j+1}\bra{{\rm e},j+1}\mathcal{H}_{\rm Int}\ket{{\rm e},j+4}}{2(E_{{\rm e}(j+4)}-E_{{\rm e}j})}\left[\frac{1}{E_{{\rm e}(j+1)}-E_{{\rm e}j}}+\frac{1}{E_{{\rm e}(j+1)}-E_{{\rm e}(j+4)}}\right]\nonumber\\
&&+\frac{\bra{{\rm e},j}\mathcal{H}_{\rm Int}\ket{{\rm g},j+3}\bra{{\rm g},j+3}\mathcal{H}_{\rm Int}\ket{{\rm e},j+4}}{2(E_{{\rm e}(j+4)}-E_{{\rm e}j})}\left[\frac{1}{E_{{\rm g}(j+3)}-E_{{\rm e}j}}+\frac{1}{E_{{\rm g}(j+3)}-E_{{\rm e}(j+4)}}\right]\nonumber\\
&&+\frac{\bra{{\rm e},j}\mathcal{H}_{\rm Int}\ket{{\rm e},j+3}\bra{{\rm e},j+3}\mathcal{H}_{\rm Int}\ket{{\rm e},j+4}}{2(E_{{\rm e}(j+4)}-E_{{\rm e}j})}\left[\frac{1}{E_{{\rm e}(j+3)}-E_{{\rm e}j}}+\frac{1}{E_{{\rm e}(j+3)}-E_{{\rm e}(j+4)}}\right]\nonumber\\
&&+\frac{\bra{{\rm e},j}\mathcal{H}_{\rm Int}\ket{{\rm g},j+1}\bra{{\rm g},j+1}\mathcal{H}_{\rm Int}\ket{{\rm e},j+4}}{2(E_{{\rm e}(j+4)}-E_{{\rm e}j})(E_{{\rm g}(j+1)}-E_{{\rm e}(j+4)})}\nonumber\\
&=&-\frac{ g^2 \sqrt{(j+1)(j+2)(j+3)(j+4)} \alpha  \Delta_0^2 \left(\Delta_b ^2-3 \Omega^2\right)}{8 \hbar \Delta_b ^2 \Omega^2 \left(\Delta_b ^3+\Omega \Delta_b ^2-9 \Omega^2 \Delta_b -9 \Omega^3\right)}+\mathcal{O}(\alpha^2),\nonumber
\end{eqnarray}

\section{Oscillator matrix elements}\label{appME}
Here we give the explicit form of the functions $L_{LO}$ and $L_{NO}$ introduced in section \ref{secME} and derive the corresponding matrix elements $y_{nm}$. The zeroth order contributions in the nonlinearity in section \ref{secME} are denoted by:
\begin{eqnarray}\label{gl22}
L_{LO1}(g^2)=\frac{g^2\Delta_0^2(2\Delta_b +3\Omega)}{2\Omega\Delta_b^2(\Omega+\Delta_b)^2},&\quad\quad\quad&
L_{LO0+}(g)=\frac{g\Delta_0}{\Delta_b(\Omega+\Delta_b)},\nonumber\\
L_{LO1+}(g^2)=\frac{4g^2\varepsilon\Delta_0}{\Delta_b^2(\Delta_b^2+3\Omega\Delta_b+2\Omega^2)},&\quad\quad\quad&
L_{LO1-}(g^2)=-\frac{4g^2\varepsilon\Delta_0}{\Delta_b^2\Omega(\Delta_b-2\Omega)}\nonumber.
\end{eqnarray}
The term independent of $g$ is $L_{NO}(\alpha)=-3\alpha/2\hbar\Omega.$\\
The terms linear in $\alpha$ and $g$ are given by:
\begin{eqnarray}\label{gl21}
L_{NO0+}(\alpha,g)&=&-\frac{3\alpha g\Delta_0(\Delta_b+2\Omega)}{\hbar\Delta_b\Omega(\Delta_b+\Omega)^2},\nonumber\\
L_{NO2+}(\alpha,g)&=&\frac{3\alpha g}{4}\frac{\Delta_0(\Delta_b^2+6\Delta_b\Omega+13\Omega^2)}{\hbar\Omega(\Delta_b+\Omega)^2(\Delta_b^2+3\Delta_b\Omega)},\nonumber\\
L_{NO2-}(\alpha,g)&=&-\frac{3\alpha g\Delta_0}{\hbar\Delta_b(\Delta_b-3\Omega)(\Delta_b+\Omega)},\nonumber\\
L_{NO2}(\alpha,g)&=&-\frac{4\alpha\varepsilon}{\hbar\Delta_b\Omega}.\nonumber
\end{eqnarray}
Finally, the terms linear in $\alpha$ but quadratic in $g$ are:
\begin{eqnarray}
L_{NO1{\rm g}}(j,\alpha,g^2)&=&-\frac{6\varepsilon^2\alpha g^2}{\hbar\Delta_b^2\Omega^3}-\frac{3\alpha g^2\Delta_0^2[14(j+1)\Delta_b^3-\Omega^2\Delta_b(88+92j)-(3+5j)\Omega\Delta_b^2-(89j+87)\Omega^3]}{4\hbar\Omega^2\Delta_b^2(\Omega+\Delta_b)^3(\Delta_b-3\Omega)},\nonumber\\
L_{NO1{\rm e}}(j,\alpha,g^2)&=&-\frac{6\varepsilon^2\alpha g^2}{\hbar\Delta_b^2\Omega^3}-\frac{3\alpha g^2\Delta_0^2[-14(j+1)\Delta_b^3+(5j+7)\Delta_b^2\Omega+\Omega^2\Delta_b(92j+96)+\Omega^3(89j+91)]}{4\hbar\Delta_b^2\Omega^2(\Delta_b-3\Omega)(\Delta_b+\Omega)^3},\nonumber\\
L_{NO1+}(\alpha,g^2)&=&-\frac{2\alpha g^2\Delta_0\varepsilon(4\Delta_b^4+29\Omega\Delta_b^3+51\Omega^2\Delta_b^2-80\Delta_b\Omega^3-124\Omega^4)}{\hbar\Omega^2(\Delta_b-2\Omega)(\Delta_b^3+3\Delta_b^2\Omega+2\Delta_b\Omega^2)^2},\nonumber\\
L_{NO1-}(\alpha,g^2)&=&+\frac{6\alpha g^2\varepsilon\Delta_0(9\Delta_b^3+\Delta_b^2\Omega-56\Omega^2\Delta_b-36\Delta_b^3)}{\hbar\Omega^2(\Delta_b^2+3\Omega\Delta_b+2\Omega^2)(\Delta_b^2-2\Delta_b\Omega)^2} \nonumber,\\
L_{NO3}(\alpha,g^2)&=&\frac{\alpha g^2\Delta_0^2(14\Delta_b^3+25\Delta_b^2\Omega-130\Omega^2\Delta_b-261\Omega^3)}{8\hbar\Delta_b^2\Omega^2(\Delta_b+\Omega)^2(\Delta_b^2-9\Omega^2)},\nonumber\\
L_{NO3+}(\alpha,g^2)&=&\frac{\alpha g^2\Delta_0\varepsilon(\Delta_b^3+3\Omega\Delta_b^2+74\Omega^2\Delta_b+216\Omega^3)}{3\hbar\Delta_b^2\Omega(\Delta_b+2\Omega)^2(\Delta_b^3+8\Delta_b^2\Omega+19\Omega^2\Delta_b+12\Omega^3)},\nonumber\\
L_{NO3-}(\alpha,g^2)&=&-\frac{\alpha g^2\Delta_0\varepsilon(24\Delta_b^3-239\Omega\Delta_b^2+814\Omega^2\Delta_b-936\Omega^3)}{3\hbar\Delta_b^2\Omega^2(\Delta_b-2\Omega)^2(\Delta_b^2-7\Delta_b\Omega+12\Omega^2)}\nonumber.
\end{eqnarray}
We can now give the expressions for $y_{nm}$ using $\eta_j=\frac{2|\Delta(j)|}{\delta_j}$, where $0\leq\eta_j<\pi$:
\begin{eqnarray}\label{ynm1}
y_{2j+1,2j+1}&=&-L_{NO0}(j+1,\alpha,g)+L_{NO0}(j,\alpha,g)-\cos\eta_j[2L_{LO0}(g)+L_{NO0}(j+1,\alpha,g)+L_{NO0}(j,\alpha,g)]\\&&
+\sqrt{j+1}\sin\eta_j\left[L_{LO1-}(g^2)+(j+1)L_{NO1-}(\alpha,g^2)\right],\nonumber\\
y_{2j+1,2j+2}&=&[2L_{LO0}(g)+L_{NO0}(j,\alpha,g)+L_{NO0}(j+1,\alpha,g)]\sin\eta_j\nonumber\\
&&+\sqrt{j+1}\cos\eta_j\left[L_{LO1-}(g^2)+(j+1)L_{NO1-}(\alpha,g^2)\right],\nonumber\\
y_{2j+1,2j+3}&=&\cos\frac{\eta_j}{2}\cos\frac{\eta_{j+1}}{2}\sqrt{j+2}\left[1+(j+2)L_{NO}(\alpha)+L_{LO1}(g^2)+L_{NO1{\rm g}}(j+1,\alpha,g^2)\right]
,\nonumber\\
&&+\cos\frac{\eta_j}{2}\sin\frac{\eta_{j+1}}{2}\left[L_{LO0+}(g)+L_{NO0+}(\alpha,g)(2(j+1)+1)\right]\nonumber\\
&&+\sin\frac{\eta_j}{2}\cos\frac{\eta_{j+1}}{2}\sqrt{(j+1)(j+2)}L_{NO2-}(\alpha,g)\nonumber\\
&&+\sin\frac{\eta_j}{2}\sin\frac{\eta_{j+1}}{2}\sqrt{j+1}\left[1+(j+1)L_{NO}(\alpha)-L_{LO1}(g^2)
+L_{NO1{\rm e}}(j,\alpha,g^2)\right],\nonumber\\
y_{2j+1,2j+4}&=&\cos\frac{\eta_j}{2}\cos\frac{\eta_{j+1}}{2}\left[L_{LO0+}(g)+L_{NO0+}(\alpha,g)(2(j+1)+1)\right]\nonumber\\
&&-\cos\frac{\eta_j}{2}\sin\frac{\eta_{j+1}}{2}\sqrt{j+2}\left[1+(j+2)L_{NO}(\alpha)+L_{LO1}(g^2)+L_{NO1{\rm g}}(j+1,\alpha,g^2)\right]\nonumber\\
&&+\sin\frac{\eta_j}{2}\cos\frac{\eta_{j+1}}{2}\left[1+(j+1)L_{NO}(\alpha)-L_{LO1}(g^2)+L_{NO1{\rm e}}(j,\alpha,g^2)\right]\nonumber\\
&&-\sin\frac{\eta_j}{2}\sin\frac{\eta_{j+1}}{2}\sqrt{(j+1)(j+2)}L_{NO2-}(\alpha,g),\nonumber\\
y_{2j+1,2j+5}&=&\cos\frac{\eta_j}{2}\cos\frac{\eta_{j+2}}{2}\sqrt{(j+2)(j+3)}L_{NO2}(\alpha,g)\nonumber\\
&&
+\cos\frac{\eta_j}{2}\sin\frac{\eta_{j+2}}{2}\sqrt{j+2}\left[L_{LO1+}(g^2)+(j+2)L_{NO1+}(\alpha,g^2)\right]\nonumber\\
&&+\sin\frac{\eta_j}{2}\cos\frac{\eta_{j+2}}{2}\sqrt{(j+1)(j+2)(j+3)}L_{NO3-}(\alpha, g^2)\nonumber\\%\\
&&-\sin\frac{\eta_j}{2}\sin\frac{\eta_{j+2}}{2}L_{NO2}(\alpha,g)\sqrt{(j+1)(j+2)},\nonumber\\
y_{2j+1,2j+6}&=&-\cos\frac{\eta_j}{2}\sin\frac{\eta_{j+2}}{2}\sqrt{(j+2)(j+3)}L_{NO2}(\alpha,g)\nonumber\\
&&
+\cos\frac{\eta_j}{2}\cos\frac{\eta_{j+2}}{2}\sqrt{j+2}\left[L_{LO1+}(g^2)+(j+2)L_{NO1+}(\alpha,g^2)\right]\nonumber\\
&&-\sin\frac{\eta_j}{2}\sin\frac{\eta_{j+2}}{2}\sqrt{(j+1)(j+2)(j+3)}L_{NO3-}(\alpha, g^2)\nonumber\\
&&-\sin\frac{\eta_j}{2}\cos\frac{\eta_{j+2}}{2}L_{NO2}(\alpha,g)\sqrt{(j+1)(j+2)}\nonumber,\\
y_{2j+1,2j+7}&=&+\cos\frac{\eta_j}{2}\cos\frac{\eta_{j+3}}{2}\sqrt{(j+2)(j+3)(j+4)}\left[L_{NO3}(\alpha,g^2)-L_{NO}(\alpha)/2\right]\nonumber\\&&+
\cos\frac{\eta_j}{2}\sin\frac{\eta_{j+3}}{2}\sqrt{(j+2)(j+3)}L_{NO2+}(\alpha,g)\nonumber\\&&+
\sin\frac{\eta_j}{2}\sin\frac{\eta_{j+3}}{2}\sqrt{(j+1)(j+2)(j+3)}\left[-L_{NO3}(\alpha,g^2)-L_{NO}(\alpha)/2\right]\nonumber,\\
y_{2j+1,2j+8}&=&-\cos\frac{\eta_j}{2}\sin\frac{\eta_{j+3}}{2}\sqrt{(j+2)(j+3)(j+4)}\left[L_{NO3}(\alpha,g^2)-L_{NO}(\alpha)/2\right]\nonumber\\&&+
\cos\frac{\eta_j}{2}\cos\frac{\eta_{j+3}}{2}\sqrt{(j+2)(j+3)}L_{NO2+}(\alpha,g)\nonumber\\&&+
\sin\frac{\eta_j}{2}\cos\frac{\eta_{j+3}}{2}\sqrt{(j+1)(j+2)(j+3)}\left[-L_{NO3}(\alpha,g^2)-L_{NO}(\alpha)/2\right]\nonumber,\\
y_{2j+1,2j+9}&=&\cos\frac{\eta_j}{2}\sin\frac{\eta_{j+4}}{2}\sqrt{(j+2)(j+3)(j+4)}L_{NO3+}(\alpha,g^2)\nonumber,\\
y_{2j+1,2j+10}&=&\cos\frac{\eta_j}{2}\cos\frac{\eta_{j+4}}{2}\sqrt{(j+2)(j+3)(j+4)}L_{NO3+}(\alpha,g^2)\nonumber,
\end{eqnarray}
and
\begin{eqnarray}
y_{2j+2,2j+2}&=&L_{NO0}(j,\alpha,g)-L_{NO0}(j+1,\alpha,g)+(2L_{LO0}(g)+L_{NO0}(j,\alpha,g)+L_{NO0}(j+1,\alpha,g))\cos\eta_j\\&&-\sin\eta_j\sqrt{j+1}\left[L_{LO1-}(g^2)+(j+1)L_{NO1-}(\alpha,g^2)\right],\nonumber\\
y_{2j+2,2j+3}&=&-\sin\frac{\eta_j}{2}\cos\frac{\eta_{j+1}}{2}\sqrt{j+2}\left[1+(j+2)L_{NO}(\alpha)+L_{LO1}(g^2)+L_{NO1{\rm g}}(j+1,\alpha,g^2)\right]
,\nonumber\\
&&-\sin\frac{\eta_j}{2}\sin\frac{\eta_{j+1}}{2}\left[L_{LO0+}(g)+L_{NO0+}(\alpha,g)(2(j+1)+1)\right]\nonumber\\
&&+\cos\frac{\eta_j}{2}\cos\frac{\eta_{j+1}}{2}\sqrt{(j+1)(j+2)}L_{NO2-}(\alpha,g)\nonumber\\
&&+\cos\frac{\eta_j}{2}\sin\frac{\eta_{j+1}}{2}\sqrt{j+1}\left[1+(j+1)L_{NO}(\alpha)-L_{LO1}(g^2)+L_{NO1{\rm e}}(j,\alpha,g^2)\right],\nonumber\\
y_{2j+2,2j+4}&=&
\sin\frac{\eta_j}{2}\sin\frac{\eta_{j+1}}{2}\sqrt{j+2}\left[1+(j+2)L_{NO}(\alpha)+L_{LO1}(g^2)+L_{NO1{\rm g}}(j+1,\alpha,g^2)\right]
\nonumber\\
&&-\sin\frac{\eta_j}{2}\cos\frac{\eta_{j+1}}{2}\left[L_{LO0+}(g)+L_{NO0+}(\alpha,g)(2(j+1)+1)\right]\nonumber\\
&&-\cos\frac{\eta_j}{2}\sin\frac{\eta_{j+1}}{2}\sqrt{(j+1)(j+2)}L_{NO2-}(\alpha,g)\nonumber\\
&&+\cos\frac{\eta_j}{2}\cos\frac{\eta_{j+1}}{2}\sqrt{j+1}\left[1+(j+1)L_{NO}(\alpha)-L_{LO1}(g^2)+L_{NO1{\rm e}}(j,\alpha,g^2)\right],\nonumber\\
y_{2j+2,2j+5}&=&
-\sin\frac{\eta_j}{2}\cos\frac{\eta_{j+2}}{2}\sqrt{(j+2)(j+3)}L_{NO2}(\alpha,g)\nonumber\\
&&
-\sin\frac{\eta_j}{2}\sin\frac{\eta_{j+2}}{2}\sqrt{j+2}\left[L_{LO1+}(g^2)+L_{NO1+}(\alpha,g^2)(j+2)\right]\nonumber\\
&&+\cos\frac{\eta_j}{2}\cos\frac{\eta_{j+2}}{2}\sqrt{(j+1)(j+2)(j+3)}L_{NO3-}(\alpha,g^2)\nonumber\\
&&-\cos\frac{\eta_j}{2}\sin\frac{\eta_{j+2}}{2}\sqrt{(j+2)(j+3)}L_{NO2}(\alpha,g),\nonumber\\
y_{2j+2,2j+6}&=&
\sin\frac{\eta_j}{2}\sin\frac{\eta_{j+2}}{2}\sqrt{(j+2)(j+3)}L_{NO2}(\alpha,g)\nonumber\\
&&
-\sin\frac{\eta_j}{2}\cos\frac{\eta_{j+2}}{2}\sqrt{j+2}\left[L_{LO1+}(g^2)+L_{NO1+}(\alpha,g^2)(j+2)\right]\nonumber\\
&&-\cos\frac{\eta_j}{2}\sin\frac{\eta_{j+2}}{2}\sqrt{(j+1)(j+2)(j+3)}L_{NO3-}(\alpha,g^2)\nonumber\\
&&-\cos\frac{\eta_j}{2}\cos\frac{\eta_{j+2}}{2}\sqrt{(j+2)(j+3)}L_{NO2}(\alpha,g)\nonumber,\\
y_{2j+2,2j+7}&=&-\sin\frac{\eta_j}{2}\cos\frac{\eta_{j+3}}{2}\sqrt{(j+2)(j+3)(j+4)}\left[L_{NO3}(\alpha,g^2)-L_{NO}(\alpha)/2\right]\nonumber\\&&-
\sin\frac{\eta_j}{2}\sin\frac{\eta_{j+3}}{2}\sqrt{(j+2)(j+3)}L_{NO2+}(\alpha,g)\nonumber\\&&+
\cos\frac{\eta_j}{2}\sin\frac{\eta_{j+3}}{2}\sqrt{(j+1)(j+2)(j+3)}\left[-L_{NO3}(\alpha,g^2)-L_{NO}(\alpha)/2\right],\nonumber\\
y_{2j+2,2j+8}&=&+\sin\frac{\eta_j}{2}\sin\frac{\eta_{j+3}}{2}\sqrt{(j+2)(j+3)(j+4)}\left[L_{NO3}(\alpha,g^2)-L_{NO}(\alpha)/2\right]\nonumber\\&&-
\sin\frac{\eta_j}{2}\cos\frac{\eta_{j+3}}{2}\sqrt{(j+2)(j+3)}L_{NO2+}(\alpha,g)\nonumber\\&&+
\cos\frac{\eta_j}{2}\cos\frac{\eta_{j+3}}{2}\sqrt{(j+1)(j+2)(j+3)}\left[-L_{NO3}(\alpha,g^2)-L_{NO}(\alpha)/2\right],\nonumber\\
y_{2j+2,2j+9}&=&-\sin\frac{\eta_j}{2}\sin\frac{\eta_{j+4}}{2}\sqrt{(j+2)(j+3)(j+4)}L_{NO3+}(\alpha,g^2),\nonumber\\
y_{2j+2,2j+10}&=&-\sin\frac{\eta_j}{2}\cos\frac{\eta_{j+4}}{2}\sqrt{(j+2)(j+3)(j+4)}L_{NO3+}(\alpha,g^2)\nonumber.
\end{eqnarray}
The matrix elements including the ground state are calculated separately because of its special form:
\begin{eqnarray}
y_{00}&=&-2(L_{LO0}(g)+L_{NO0}(0,\alpha,g)),\\
y_{01}&=&\cos\frac{\eta_0}{2}\left[1+L_{NO}(\alpha)+L_{LO1}(g^2)+L_{NO1{\rm g}}(0,\alpha,g^2)\right]+\sin\frac{\eta_0}{2}\left[L_{LO0+}(g)+L_{NO0+}(\alpha,g)\right]\nonumber,\\
y_{02}&=&-\sin\frac{\eta_0}{2}\left[1+L_{NO}(\alpha)+L_{LO1}(g^2)+L_{NO1{\rm g}}(0,\alpha,g^2)\right]+\cos\frac{\eta_0}{2}\left[L_{LO0+}(g)+L_{NO0+}(\alpha,g)\right]\nonumber,\\
y_{03}&=&\cos\frac{\eta_1}{2}\sqrt{2}L_{NO2}(\alpha,g)+\sin\frac{\eta_1}{2}\left[L_{LO1+}(g^2)+L_{NO1+}(\alpha,g^2)\right]\nonumber,\\
y_{04}&=&-\sin\frac{\eta_1}{2}\sqrt{2}L_{NO2}(\alpha,g)+\cos\frac{\eta_1}{2}\left[L_{LO1+}(g^2)+L_{NO1+}(\alpha,g^2)\right]\nonumber,\\
y_{05}&=&\cos\frac{\eta_2}{2}\sqrt{3}\left[L_{NO3}(\alpha,g^2)-L_{NO}(\alpha)/2\right]+\sin\frac{\eta_2}{2}\sqrt{2}L_{NO2+}(\alpha,g)\nonumber,\\
y_{06}&=&-\sin\frac{\eta_2}{2}\sqrt{3}\left[L_{NO3}(\alpha,g^2)-L_{NO}(\alpha)/2\right]+\cos\frac{\eta_2}{2}\sqrt{2}L_{NO2+}(\alpha,g)\nonumber,\\
y_{07}&=&\sin\frac{\eta_3}{2}\sqrt{3}L_{NO3+}(\alpha,g^2)\nonumber,\\
y_{08}&=&\cos\frac{\eta_3}{2}\sqrt{3}L_{NO3+}(\alpha,g^2)\nonumber.
\end{eqnarray}
\section{Rate coefficients for the off-diagonal density matrix elements}\label{appdeph}
We give the rate coefficients occurring in the Bloch-Redfield equation (\ref{gl13}) for the reduced density matrix,
\begin{equation}
  \mathcal{L}_{01,01} = \frac{2 \kappa}{\hbar \beta}y_{00}y_{11}-\frac{ \kappa}{\hbar \beta}y_{00}^2-\frac{\kappa}{\hbar \beta}y_{11}^2 - \frac{1}{2} \mathcal{L}_{00,11}, \label{L0101}
\end{equation} 
\begin{equation}
  \mathcal{L}_{02,02} =\frac{2 \kappa}{\hbar \beta}y_{00}y_{22}-\frac{ \kappa}{\hbar \beta}y_{00}^2-\frac{\kappa}{\hbar \beta}y_{22}^2 - \frac{1}{2} \mathcal{L}_{00,22}, \label{L0202}
\end{equation}
\begin{equation}
   \mathcal{L}_{03,03} =\frac{2 \kappa}{\hbar \beta}y_{00}y_{33}-\frac{ \kappa}{\hbar \beta}y_{00}^2-\frac{\kappa}{\hbar \beta}y_{33}^2- \frac{1}{2} \mathcal{L}_{11,33}- \frac{1}{2} \mathcal{L}_{22,33},
\end{equation}
\begin{equation}
  \mathcal{L}_{04,04} =\frac{2 \kappa}{\hbar \beta}y_{00}y_{44}-\frac{ \kappa}{\hbar \beta}y_{00}^2-\frac{\kappa}{\hbar \beta}y_{44}^2-\frac{1}{2} \mathcal{L}_{11,44}- \frac{1}{2} \mathcal{L}_{22,44},
\end{equation}
\begin{equation}
    \mathcal{L}_{12,12} =\frac{2 \kappa}{\hbar \beta}y_{11}y_{22}-\frac{ \kappa}{\hbar \beta}y_{11}^2-\frac{\kappa}{\hbar \beta}y_{22}^2 - \frac{1}{2} \mathcal{L}_{00,11}- \frac{1}{2} \mathcal{L}_{00,22},
\end{equation} 
\begin{equation}
   \mathcal{L}_{13,13} =\frac{2 \kappa}{\hbar \beta}y_{11}y_{33}-\frac{ \kappa}{\hbar \beta}y_{11}^2-\frac{\kappa}{\hbar \beta}y_{33}^2 - \frac{1}{2} \mathcal{L}_{00,11}- \frac{1}{2} \mathcal{L}_{11,33} - \frac{1}{2} \mathcal{L}_{22,33},
\end{equation}
\begin{equation}
 \mathcal{L}_{14,14} =\frac{2 \kappa}{\hbar \beta}y_{11}y_{44}-\frac{ \kappa}{\hbar \beta}y_{11}^2-\frac{\kappa}{\hbar \beta}y_{44}^2  - \frac{1}{2} \mathcal{L}_{00,11}- \frac{1}{2} \mathcal{L}_{11,44} - \frac{1}{2} \mathcal{L}_{22,44},
\end{equation}
\begin{equation}
   \mathcal{L}_{23,23} =\frac{2 \kappa}{\hbar \beta}y_{22}y_{33}-\frac{ \kappa}{\hbar \beta}y_{22}^2-\frac{\kappa}{\hbar \beta}y_{33}^2- \frac{1}{2} \mathcal{L}_{00,22}- \frac{1}{2} \mathcal{L}_{11,33} - \frac{1}{2} \mathcal{L}_{22,33},
\end{equation}
\begin{equation}
  \mathcal{L}_{24,24} =\frac{2 \kappa}{\hbar \beta}y_{22}y_{44}-\frac{ \kappa}{\hbar \beta}y_{22}^2-\frac{\kappa}{\hbar \beta}y_{44}^2 - \frac{1}{2} \mathcal{L}_{00,22}- \frac{1}{2} \mathcal{L}_{11,44} - \frac{1}{2} \mathcal{L}_{22,44},
\end{equation}
\begin{equation}
  \mathcal{L}_{34,34} = \frac{2 \kappa}{\hbar \beta}y_{33}y_{44}-\frac{ \kappa}{\hbar \beta}y_{33}^2-\frac{\kappa}{\hbar \beta}y_{44}^2 - \frac{1}{2} \mathcal{L}_{11,33} - \frac{1}{2} \mathcal{L}_{22,33} - \frac{1}{2} \mathcal{L}_{11,44} - \frac{1}{2} \mathcal{L}_{22,44},
\end{equation}
\begin{eqnarray}
  \mathcal{L}_{01,02} &=& \frac{ \kappa}{\hbar \beta} (y_{00}y_{12} - y_{12}y_{22}) - G(\omega_{02}) N_{02} y_{01} y_{02} - G(\omega_{12}) N_{12} y_{11} y_{12} \nonumber\\
  && - G(\omega_{32}) N_{32} y_{13} y_{23} - G(\omega_{42}) N_{42} y_{14} y_{24}, 
\end{eqnarray}
\begin{eqnarray}
  \mathcal{L}_{02,01} &=& \frac{ \kappa}{\hbar \beta} (y_{00}y_{12} - y_{12}y_{11}) - G(\omega_{01}) N_{01} y_{01} y_{02} - G(\omega_{21}) N_{21} y_{22} y_{12} \nonumber\\
   &&- G(\omega_{31}) N_{31} y_{13} y_{23} - G(\omega_{41}) N_{41} y_{14} y_{24}, 
\end{eqnarray}
\begin{eqnarray}
  \mathcal{L}_{13,23} &=& \frac{ \kappa}{\hbar \beta} (y_{33}y_{12} - y_{12}y_{22}) - G(\omega_{12}) N_{12} (y_{11} y_{12} - y_{12} y_{33}) \nonumber\\
    &&- G(\omega_{02}) N_{02} y_{01} y_{02} - G(\omega_{32}) N_{32} y_{13} y_{23} - G(\omega_{42}) N_{42} y_{14} y_{24}, 
\end{eqnarray}
\begin{eqnarray}
  \mathcal{L}_{23,13} &=& \frac{ \kappa}{\hbar \beta} (y_{33}y_{12} - y_{12}y_{11}) - G(\omega_{21}) N_{21} (y_{22} y_{12} - y_{12} y_{33}) \nonumber\\
   &&- G(\omega_{01}) N_{01} y_{01} y_{02} - G(\omega_{31}) N_{31} y_{13} y_{23} - G(\omega_{41}) N_{41} y_{14} y_{24}, 
\end{eqnarray}
\begin{eqnarray}
   \mathcal{L}_{14,24} &=& \frac{ \kappa}{\hbar \beta} (y_{44}y_{12} - y_{12}y_{22}) - G(\omega_{12}) N_{12} (y_{11} y_{12} - y_{12} y_{44}) \nonumber\\
  &&  - G(\omega_{02}) N_{02} y_{01} y_{02} - G(\omega_{32}) N_{32} y_{13} y_{23} - G(\omega_{42}) N_{42} y_{14} y_{24}, 
\end{eqnarray}
\begin{eqnarray}
  \mathcal{L}_{24,14} &=& \frac{ \kappa}{\hbar \beta} (y_{44}y_{12} - y_{12}y_{11}) - G(\omega_{21}) N_{21} (y_{22} y_{12} - y_{12} y_{44}) \nonumber\\
  && - G(\omega_{01}) N_{01} y_{01} y_{02} - G(\omega_{31}) N_{31} y_{13} y_{23} - G(\omega_{41}) N_{41} y_{14} y_{24}. 
\end{eqnarray}
\section{Diagonal reduced density matrix elements}\label{AppDiagonalElements}
The solutions of the FSA master equation (\ref{gl14a}) for the diagonal elements within the low temperature approximation (\ref{SimplifiedMasterEq}) reads:
\begin{eqnarray} 
  \sigma_{00}(t) &=& \sigma_{00}^0  + \sigma_{11}^0 + \sigma_{22}^0 + \sigma_{33}^0+ \sigma_{44}^0\nonumber \\ &&
  - \exp(- \pi \mathcal{L}_{00,11} t) \biggl( \sigma_{11}^0 + \sigma_{33}^0 \frac{\mathcal{L}_{11,33}}{-\mathcal{L}_{00,11} + \mathcal{L}_{11,33} + \mathcal{L}_{22,33}} 
     + \sigma_{44}^0 \frac{\mathcal{L}_{11,44}}{-\mathcal{L}_{00,11} + \mathcal{L}_{11,44} + \mathcal{L}_{22,44}} \biggr) \nonumber \\ &&
   - \exp(- \pi \mathcal{L}_{00,22} t) \biggl( \sigma_{22}^0 + \sigma_{33}^0 \frac{\mathcal{L}_{22,33}}{-\mathcal{L}_{00,22} + \mathcal{L}_{11,33} + \mathcal{L}_{22,33}}  
                + \sigma_{44}^0 \frac{\mathcal{L}_{22,44}}{-\mathcal{L}_{00,22} + \mathcal{L}_{11,44} + \mathcal{L}_{22,44}} \biggr) \nonumber \\  &&
  + \exp(- \pi ( \mathcal{L}_{11,33} + \mathcal{L}_{22,33}) t)  \sigma_{33}^0 \biggl(  \frac{ \mathcal{L}_{00,22} - \mathcal{L}_{11,33}  }{-\mathcal{L}_{00,22} + \mathcal{L}_{11,33} + \mathcal{L}_{22,33}} 
            + \frac{ \mathcal{L}_{11,33}  }{-\mathcal{L}_{00,11} + \mathcal{L}_{11,33} + \mathcal{L}_{22,33}}  \biggr) \nonumber\\ &&
   + \exp(- \pi ( \mathcal{L}_{11,44} + \mathcal{L}_{22,44}) t)  \sigma_{44}^0 \biggl(  \frac{ \mathcal{L}_{00,22} - \mathcal{L}_{11,44}  }{-\mathcal{L}_{00,22} + \mathcal{L}_{11,44} + \mathcal{L}_{22,44}} 
            + \frac{ \mathcal{L}_{11,44}  }{-\mathcal{L}_{00,11} + \mathcal{L}_{11,44} + \mathcal{L}_{22,44}}  \biggr) \label{SolDiagFSA00},
\end{eqnarray}
\begin{eqnarray}
    \sigma_{11}(t) &=& - \exp(- \pi \mathcal{L}_{00,11} t)  \sigma_{11}^0 \nonumber  \\ &&
    - \exp(- \pi ( \mathcal{L}_{00,11} + \mathcal{L}_{11,33} + \mathcal{L}_{22,33}) t)  \sigma_{33}^0  \frac{ \mathcal{L}_{11,33}  }{-\mathcal{L}_{00,11} + \mathcal{L}_{11,33} + \mathcal{L}_{22,33}} \nonumber\\  &&
     - \exp(- \pi (  \mathcal{L}_{00,11} +\mathcal{L}_{11,44} + \mathcal{L}_{22,44}) t)  \sigma_{44}^0  \frac{ \mathcal{L}_{11,44}  }{-\mathcal{L}_{00,11} + \mathcal{L}_{11,44} + \mathcal{L}_{22,44}}, \label{SolDiagFRWA11}
\end{eqnarray}
\begin{eqnarray}
  \sigma_{22}(t) &=& - \exp(- \pi \mathcal{L}_{00,22} t)  \sigma_{22}^0 \nonumber \\ &&
    - \exp(- \pi ( \mathcal{L}_{00,22} + \mathcal{L}_{11,33} + \mathcal{L}_{22,33}) t)  \sigma_{33}^0  \frac{ \mathcal{L}_{22,33}  }{-\mathcal{L}_{00,22} + \mathcal{L}_{11,33} + \mathcal{L}_{22,33}} \nonumber\\  &&
    - \exp(- \pi ( \mathcal{L}_{00,22} + \mathcal{L}_{11,44} + \mathcal{L}_{22,44}) t)  \sigma_{44}^0  \frac{ \mathcal{L}_{22,44}  }{-\mathcal{L}_{00,22} + \mathcal{L}_{11,44} + \mathcal{L}_{22,44}}, \label{SolDiagFRWA22}
\end{eqnarray}
\begin{equation}
 \sigma_{33}(t) =  \exp(- \pi ( \mathcal{L}_{11,33} + \mathcal{L}_{22,33}) t)  \sigma_{33}^0, \label{SolDiagFRWA33}
\end{equation} 
\begin{equation}
   \sigma_{44}(t) =  \exp(- \pi ( \mathcal{L}_{11,44} + \mathcal{L}_{22,44}) t)  \sigma_{44}^0. \label{SolDiagFSA44}
\end{equation}
\end{widetext}

\end{document}